\documentclass[12pt,preprint]{aastex}

\usepackage[dvips]{epsfig}
\usepackage[dvips]{rotating}

\newcommand{\he}{HE~0107$-$5240}
\newcommand{\cd}{CD~$-38^{\circ}\,245$}
\newcommand{\tefft}{$T_{\mbox{\scriptsize eff}}$}
\newcommand{\teffm}{T_{\mbox{\scriptsize eff}}}

\slugcomment{Draft version, Thu Nov  6 20:29:11 CET 2003}

\shorttitle{Abundance analysis of {\he}}
\shortauthors{Christlieb et al.}

\begin{document}

\title{{\he}, A Chemically Ancient Star.\\
  I. A Detailed Abundance Analysis\altaffilmark{1}}

\author{N. Christlieb\altaffilmark{2,3}, B. Gustafsson\altaffilmark{2},
A. J. Korn\altaffilmark{4,2}, P. S. Barklem\altaffilmark{2},
T. C. Beers\altaffilmark{5}, M. S. Bessell\altaffilmark{6},
T. Karlsson\altaffilmark{2}, M. Mizuno-Wiedner\altaffilmark{2}}

\altaffiltext{1}{Based on observations collected at the European Southern
  Observatory, Paranal, Chile (Proposal Number 268.D-5745).}
\altaffiltext{2}{Uppsala Astronomical Observatory, Box 524, SE-75239 Uppsala, Sweden}
\altaffiltext{3}{Marie Curie Fellow, on sabbatical leave from Hamburger
  Sternwarte, Gojenbergsweg 112, D-21029 Hamburg, Germany}
\altaffiltext{4}{Universit\"ats-Sternwarte M\"unchen, Scheinerstra\ss{}e 1,
  D-81679 M\"unchen, Germany}
\altaffiltext{5}{Department of Physics and Astronomy, Michigan State University, U.S.A.}
\altaffiltext{6}{Research School of Astronomy and Astrophysics, Mount Stromlo
    Observatory, Cotter Road, Weston, ACT 2611, Australia}

\begin{abstract} 
We report a detailed abundance analysis for {\he}, a halo giant with
$\mbox{[Fe/H]}_{\mbox{\scriptsize NLTE}}=-5.3$. This star was discovered in the
course of follow-up medium-resolution spectroscopy of extremely metal-poor
candidates selected from the digitized Hamburg/ESO objective-prism survey.  On
the basis of high-resolution VLT/UVES spectra, we derive abundances for 8
elements (C, N, Na, Mg, Ca, Ti, Fe, and Ni), and upper limits for another 12
elements. A plane-parallel LTE model atmosphere has been specifically tailored
for the chemical composition of {\he}. Scenarios for the origin of the abundance
pattern observed in the star are discussed. We argue that {\he} is most likely
not a post-AGB star, and that the extremely low abundances of the iron-peak, and
other elements, are not due to selective dust depletion. The abundance pattern
of {\he} can be explained by pre-enrichment from a zero-metallicity type-II
supernova of $20$--$25$\,M$_{\odot}$, plus either self-enrichment with C and N,
or production of these elements in the AGB phase of a formerly more massive
companion, which is now a white dwarf. However, significant radial velocity
variations have not been detected within the 52 days covered by our moderate-
and high-resolution spectra. Alternatively, the abundance pattern can be
explained by enrichment of the gas cloud from which {\he} formed by a
$25\mbox{M}_{\odot}$ first-generation star exploding as a subluminous SN~II, as
proposed by Umeda \& Nomoto (2003). We discuss consequences of the existence of
{\he} for low-mass star formation in extremely metal-poor environments, and for
currently ongoing and future searches for the most metal-poor stars in the
Galaxy.
\end{abstract}



\section{INTRODUCTION}

It has become clear, in recent years, that in order to understand the history of
galaxy formation, and indeed, the early evolution of the universe as a whole, it
is necessary to understand the nature of star formation at the earliest epochs.
A number of recent studies have suggested that the very first stars that formed
may have been quite massive objects, several hundred to perhaps one thousand
solar masses \citep{Brommetal:2001,Schneideretal:2002}, while others have
suggested that the First Mass Function (FMF) may have included stars with masses
as low as roughly the solar mass
\citep{Yoshii/Saio:1986,Nakamura/Umemura:2001}. The most massive stars would
have disappeared after some tens of million years, and in this process may have
contributed the first heavy element production in the early universe. The less
massive stars, especially if these included stars of sufficiently low mass that
their main-sequence lifetimes exceed a Hubble time, should still be observable
at present. If they were indeed present at the earliest times, these stars
should, to a large extent, have preserved in their atmospheres the fossil record
of the element production of the most massive stars that were their immediate
precursors.

The fundamental role that early-formed, low-mass stars play as the ``scribes''
of stellar generations from long ago has inspired a number of systematic
searches for their presence in the Galaxy today. The HK survey of Beers, Preston
and Shectman \citep{BPSII,TimTSS} was the first survey to detect significant
numbers of extremely metal-poor (EMP) stars, those which, for the purpose of
this paper, we define to be stars of metallicity\footnote{We use the common
notation $\mbox{[X/H]}=\log_{10}\mbox{[$N$(X)/$N$(H)]}_{\ast}-
\log_{10}\mbox{[$N$(X)/$N$(H)]}_{\odot}$, and analogously for [X/Fe].
Abundances are on a scale where $\log\epsilon(\mbox{H})=12$, i.e.,
$\log\epsilon(\mbox{X}) = \log\left\{N(\mbox{X})/N(\mbox{H})\right\}+12$.}
$\mbox{[Fe/H]}<-3.0$.  However, even this effort, spanning some two decades, has
to date only identified about 100 stars of such extremely low metallicity.  The
ESO Large Programme on ``Galaxy Formation, Early Nucleosynthesis, and the First
Stars'' of Cayrel et al.
\citep{Cayreletal:2001,Hilletal:2002,Depagneetal:2002,Cayreletal:2003,Francoisetal:2003}
has obtained high-resolution, high signal-to-noise ratio ($S/N$) spectra of many
of the HK survey stars with $\mbox{[Fe/H]}<-3.0$. However, it is clear that even
after this sample of stars has been analysed, there will remain many open
questions, and that new questions will arise.

In order to increase the number of known EMP stars suitable as targets for
high-resolution, high $S/N$ spectroscopy with currently existing telescopes, the
data base of digital spectra of the Hamburg/ESO objective-prism survey
\citep[HES;][]{hespaperIII} is currently being exploited by means of
quantitative selection criteria
\citep{HESstarsI,HESStarsII,MiningTheSky,HESautoclass,Christlieb:2003}. Medium-resolution
($\sim 2$\,{\AA}) spectroscopic follow-up observations of the $8\,715$ candidate
metal-poor stars identified to date (main-sequence turnoff stars and subgiants
as well as giants) are being obtained with 1.5m--6.5\,m telescopes at AAO, CTIO,
ESO, KPNO, LCO, Palomar Observatory and SSO. To date, follow-up spectra of $\sim
3\,300$ HES metal-poor candidates have been obtained, and $\sim 200$ stars with
$\mbox{[Fe/H]}<-3.0$ have been identified \citep{Christlieb:2003}, which triples
the number of known EMP stars, from the $\sim 100$ found in the HK survey, to a
total of $\sim 300$ stars.  Results from high-resolution spectroscopy of HES EMP
stars obtained with VLT-UT2/UVES and Keck/HIRES have been reported in
\citet{Depagneetal:2000}, \citet{KeckpaperI}, \citet{KeckpaperII},
\citet{KeckpaperIII}, and \citet{Cohenetal:2003}.

The lowest metallicity stars in the HK survey reach to $\mbox{[Fe/H]}=-4.0$,
equal to, but not lower than, the metallicity of the lowest metallicity star
known prior to the HK survey, {\cd} \citep{Bessell/Norris:1984}. Hence, it was
widely assumed that a physical low-metallicity limit for Galactic halo stars was
reached at around $\mbox{[Fe/H]}=-4.0$.  However, in a previous paper \citep[][
hereafter Paper~I]{HE0107_Nature}, we reported the discovery of {\he}, which is
a factor of 20 times lower in [Fe/H] than {\cd}. In this paper, we describe the
derivation of the stellar parameters for {\he}, and the abundance analysis in
more detail (\S \ref{Sect:StellarParameters} and \S \ref{Sect:AbundAnalysis},
respectively).  In \S \ref{Sect:AbundancePattern} we present possible scenarios
for the origin of the abundance pattern observed in {\he}. We conclude our paper
with a discussion of some consequences of these scenarios, and consider the
impact of the existence of {\he} on theories of low-mass star formation in the
early Universe.

\section{OBSERVATIONS}

\subsection{Spectroscopy}

\begin{figure*}[htbp]
  \begin{center}
  \epsfig{file=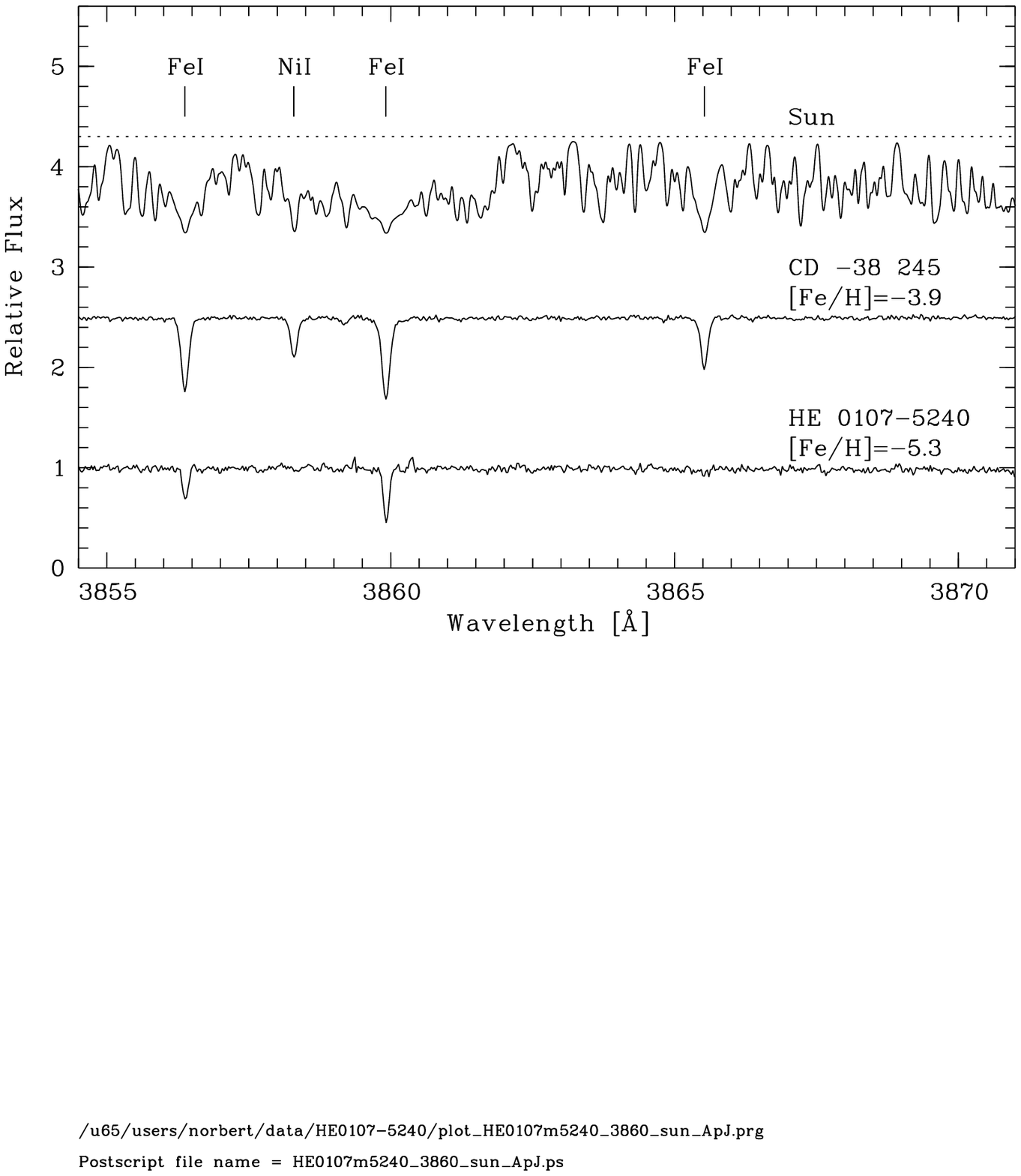, clip=, width=14cm,
    bbllx=62, bblly=335, bburx=527, bbury=627}\\[2ex]
  \epsfig{file=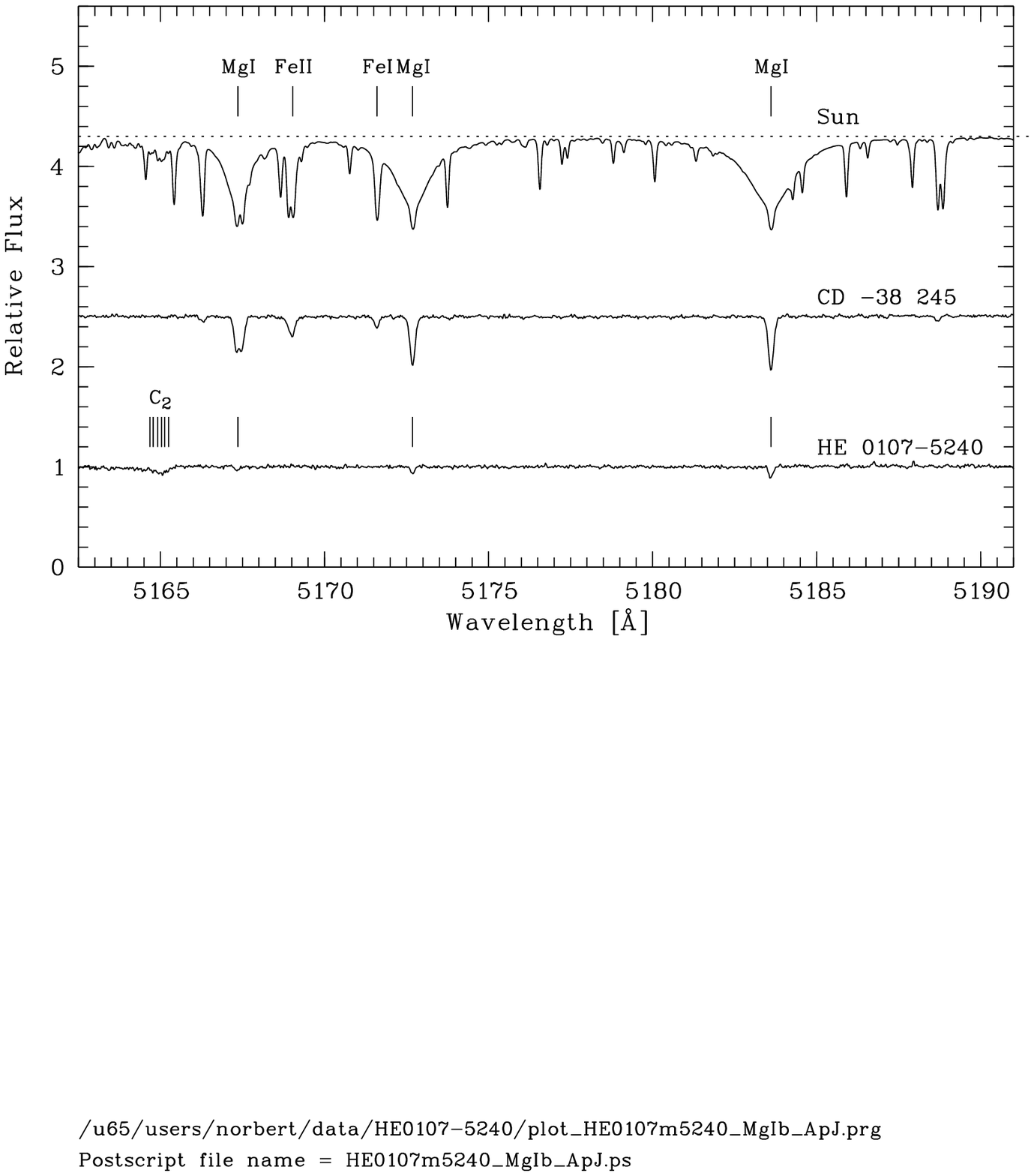, clip=, width=14cm,
    bbllx=62, bblly=335, bburx=527, bbury=627}    
  \end{center}
  \centering
  \caption{\label{Fig:SunCDm38HE0107} Spectrum of the Sun compared with the VLT/UVES
    spectrum of {\cd}, the previously most metal-poor giant star known, and with {\he}. The
    spectrum of {\cd} was obtained in our program with the same observational setup.
    The spectra are on the same scale and have been offset arbitrarily in the $y$
    direction. Note the very weak or absent Fe lines in the spectrum of {\he},
    and the presence of the C$_2$ band at $\sim 5165$\,{\AA}. }
\end{figure*}

{\he} was observed during the nights of 19 and 20 December 2001 at the European
Southern Observatory (ESO), Paranal, Chile, with the Ultraviolet-Visual
Ech{\'e}lle Spectrograph \citep[UVES;][]{Dekkeretal:2000} mounted on the 8\,m Unit
Telescope 2 (Kueyen) of the Very Large Telescope (VLT). The observations are
summarized in Table \ref{Tab:UVESobs}, and selected sections of the spectra are shown in
Figure \ref{Fig:SunCDm38HE0107}.

A dichroic beam splitter (Dichroic \#1) was used to distribute the light
collected with the telescope to the blue and red arms of UVES. The BLUE~390 and
RED~580 settings were used, yielding a wavelength coverage of 3290--4519\,{\AA}
and 4781--6807\,{\AA}, respectively, with a gap in the range
5756--5834\,{\AA}. In order to reach a resolving power of $R=40,000$, the slit
was set to a width of $1''$. The CCD binning was $1\times 1$ pixels, i.e., the
projected pixel sizes were $0.215''$ (blue arm) and $0.155''$ (red arm). The
total exposure time for {\he} was 13,800\,sec, which was split into 5 exposures
in order to facilitate removal of cosmic ray hits.

The previously most metal-poor giant star known, \cd{}
\citep{Bessell/Norris:1984}, having $\mbox{[Fe/H]}=-3.92$ \citep{Ryanetal:1996},
was observed with the same setup, on the night of 20 December 2001. The exposure
time was 900\,sec.

Throughout most of our analysis we used the pipeline-reduced UVES
spectra as provided by the ESO Data Management \& Operations Division (DMD),
except for wavelength regions that turned out to be very critical for the
analysis. In these cases, the data were re-reduced with the \texttt{REDUCE}
package of \citet{Piskunov/Valenti:2002}. \texttt{REDUCE} has been demonstrated
by the authors to be more reliable in the rejection of bad pixels than the
UVES pipeline.

The differences in geocentric radial velocity between the individual exposures
was found to be negligible (i.e., 1/3 pixel in the blue-arm spectra and 1/2
pixel in the red-arm spectra), and the reduced spectra were therefore co-added
without applying any radial velocity corrections. After that, the spectra were
rebinned by a factor of 2. The resulting spectra have an average signal-to-noise
ratio ($S/N$) per pixel of 54 (blue arm), 127 (red arm, lower CCD) and 143 (red
arm, upper CCD). The $S/N$ of the blue-arm spectra reaches a maximum of $\sim
80$ at $\lambda\sim 4200$\,{\AA}, and drops to $\sim 10$ at 3290\,{\AA}. The
$S/N$ in the lower red-arm spectrum continuously increases from $\sim 120$ to
$\sim 140$ over the covered wavelength range; in the upper red-arm spectrum it
is almost constant at $\sim 140$. The quality of the spectra of {\cd} are almost
identical, apart from the presence of a fringing pattern with an amplitude of
$\sim 2$\,\% in the red spectra which is not present in the spectra of {\he}.

For abundance analyses of EMP stars, the quality of the spectra used is very
critical, since their spectral lines are very weak. \citet[][ hereafter
NRB01]{Norrisetal:2001} give the formula
\begin{equation}
  \label{Eq:sigWlambda}
     \sigma = \frac{\lambda \cdot \sqrt{n}}{R \cdot S/N}
\end{equation}
for the $1\,\sigma$ uncertainty in the measurement of the equivalent width of a
spectral line at wavelength $\lambda$ covered by $n$ pixels of a spectrum with
resolving power $R$. They define a figure of merit, $F$, that is inversely
proportional to $\sigma$, i.e,
\begin{equation}
  \label{Eq:F}
     F = \frac{R \cdot S/N}{\lambda},
\end{equation}
where $\lambda$ is given in {\AA}, in order to compare the quality of the data
used by different authors for the analysis of metal-poor stars.

With the five stars analysed by NRB01, these authors heralded
the era of abundance analysis of EMP stars based on very high-quality ($F>500$)
spectra. They were followed recently by \citet{KeckpaperI} and
\citet{KeckpaperII}, who analysed a sample of six EMP stars (and seven
additional stars with $\mbox{[Fe/H]}<-2.5$) using Keck/HIRES spectra with
$F>600$. The \cite{Cayreletal:2003} observations of EMP star from the HK survey,
obtained with UVES, achieved figures of merit $F>850$ in their blue spectra, and
$F>650$ in their red spectra. Our co-added and rebinned UVES spectra have
$<F>=460$ (blue arm), $<F>=820$ (red arm, lower part) and $<F>=770$ (red arm,
upper part), which, according to Equation (\ref{Eq:sigWlambda}), enables us to
detect spectral lines as weak as $W_{\lambda}=10$\,m{\AA} at a $3\,\sigma$
significance level or higher in both of the red-arm spectra, and lines with
$W_{\lambda}=20$\,m{\AA} in the blue-arm spectrum at $\lambda > 3700$\,{\AA}.

In addition to the UVES spectra of {\he}, three moderate-resolution spectra are
available, obtained with the Double Beam Spectrograph (DBS) at the Siding Spring
Observatory (SSO) 2.3\,m telescope by one of us (M.S.B.). These observations
were made during the nights of 11 November and 12 December 2001, and 3 January
2002.  The discovery spectrum of November covered $3400$--$5300${\AA} at a
resolution of 2.2\,{\AA}; the later spectra covered $3620$--$4580$\,{\AA} at
1.1\,{\AA} resolution.

\begin{figure*}[htbp]
  \epsfig{file=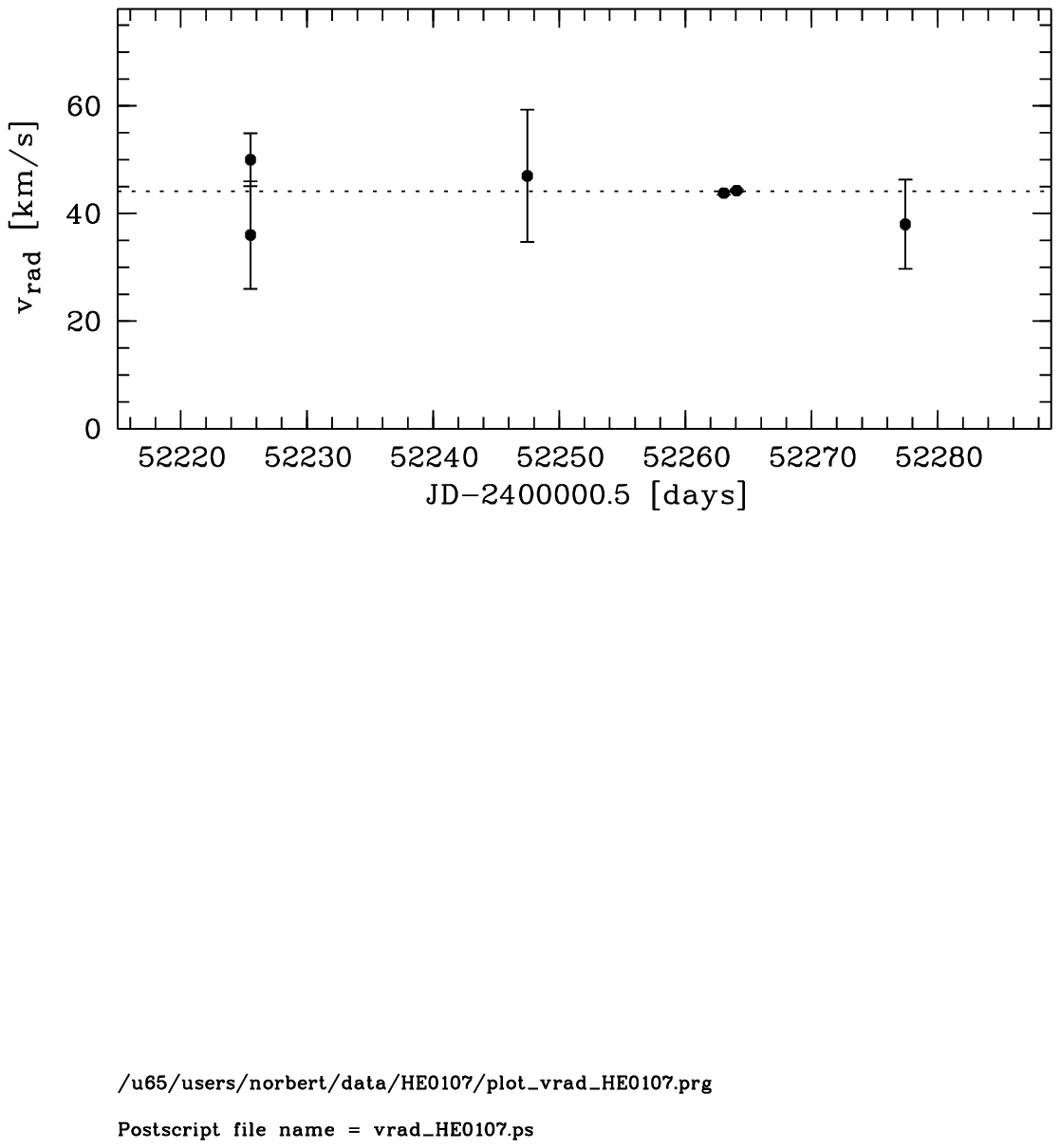, clip=, width=14cm,
    bbllx=76, bblly=284, bburx=401, bbury=443}
  \centering
  \caption{\label{Fig:vrad_HE0107} Barycentric radial velocities of {\he},
    derived from medium- and high-resolution spectra. The error bars indicate
    $1\,\sigma$ errors. The error bars for the two UVES observations are
    smaller than the symbols. The dotted line is the weighted average of all
    measurements. All measurements are consistent with a constant radial
    velocity during the period covered by the observations (52 days). The
    spectrum obtained at MJD $52225.528$ has been analysed independently by
    two persons (T.C.B. and M.S.B.), employing different methods. }
\end{figure*}

Our medium- and high-resolution spectra together cover a period of 52 days.
This set of spectra does not reveal any indication of radial velocity variations
(see Figure \ref{Fig:vrad_HE0107}). All measurements are consistent with a
constant barycentric radial velocity of $44.1\pm 0.4$ \,km\,s$^{-1}$ (with
$0.4$\,km\,s$^{-1}$ referring to the $1\,\sigma$ scatter of the measured values),
which is the weighted average of all 5 measurements we have thus far.

\subsection{Photometry}

Accurate broad- and medium-band photometry has been obtained for {\he} during
the course of several observing campaigns. The results are summarized in Table
\ref{Tab:Photometry}.

T.C.B. measured Johnson-Cousins $BVRI$ magnitudes for {\he} in December 2001
with the ESO-Danish 1.54\,m telescope and the DFOSC instrument.  Str\"omgren
$uvby$ and Johnson $V$ photometry was obtained during the night of 29 December
2001 at the Las Campanas 1\,m telescope with the Direct CCD Camera.  The
observer was M. Dehn.

$JHK$ photometry was obtained during the night of 22 December 2001 at SSO with
the 2.3\,m telescope and the CASPIR instrument by P. Wood. The measurements
were converted to the Johnson-Glass system \citep{Bessell/Brett:1988}. An
independent set of $JHK$ measurements comes from the Two Micron All Sky
Survey\footnote{2MASS is a joint project of the University of Massachusetts and
the Infrared Processing and Analysis Center/California Institute of Technology,
funded by the National Aeronautics and Space Administration and the
U.S. National Science Foundation.}  \citep[2MASS;][]{Skrutskieetal:1997}.  The
2MASS values were also transformed to the Johnson-Glass system, using the
transformations listed in \citet{Carpenter:2001}. The transformed 2MASS values
agree very well with the SSO values, to within 0.02--0.03\,mag, suggesting that
there are no systematic offsets present after the transformations were
applied. The two sets of measurements were therefore averaged to the values
listed in the last row of Table \ref{Tab:Photometry}.

The $1\,\sigma$ uncertainties of the above mentioned measurements are
typically $0.02$--$0.03$\,mag, with the exception of the Str\"omgren $c_1$
photometry. The $c_1$ magnitude has an accuracy of $0.05$\,mag. It is planned
to obtain a more accurate measurement in a future observing run.


%

\section{STELLAR PARAMETERS}\label{Sect:StellarParameters}

The adopted stellar parameters are summarized in Table
\ref{Tab:StellarParameters}. In this section we describe in detail how they have
been derived.

\subsection{Effective Temperature}\label{Sect:Teff}

\begin{figure*}[htbp]
  \epsfig{file=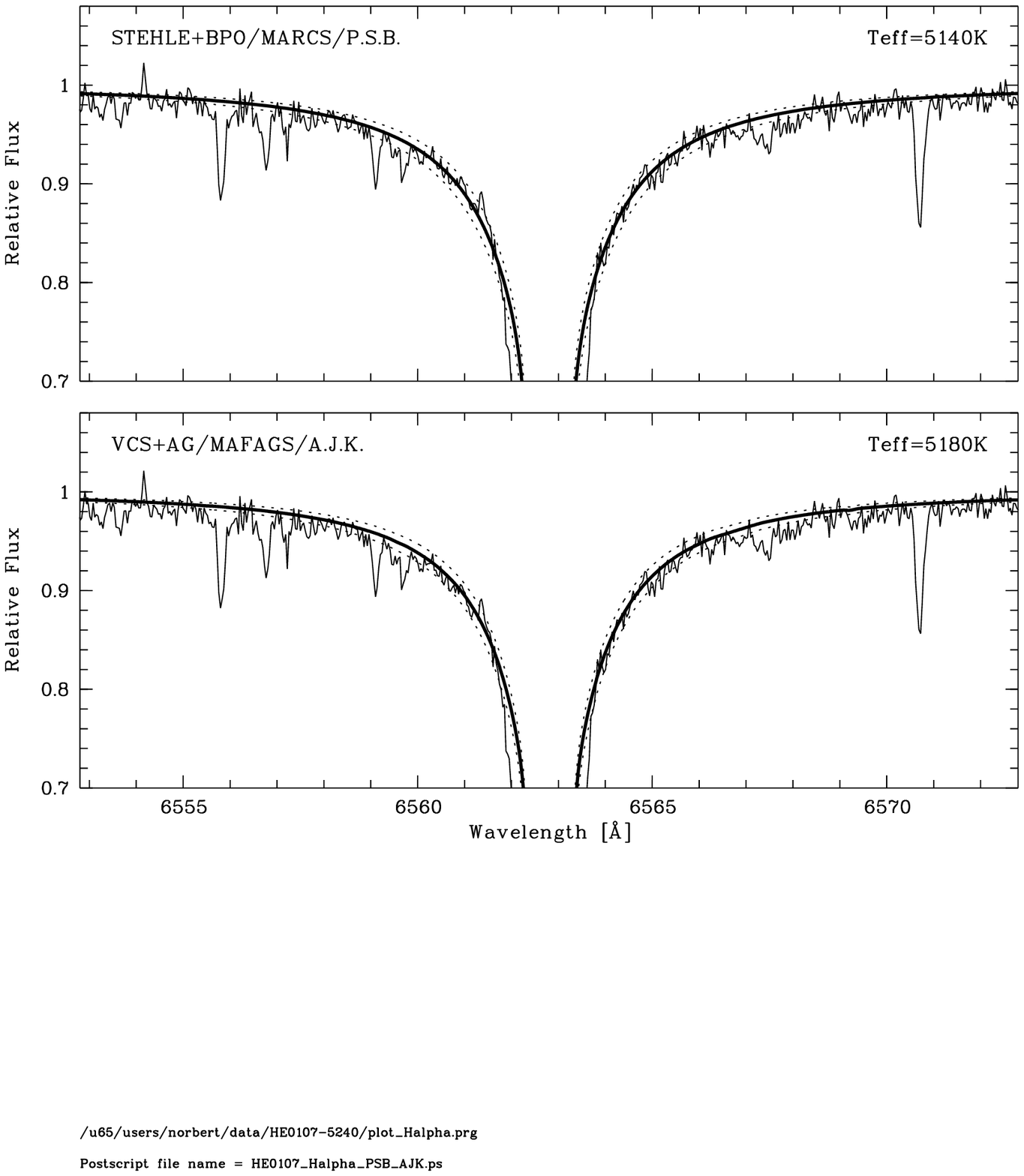, clip=, width=14cm,
    bbllx=63, bblly=241, bburx=527, bbury=629}
  \centering
  \caption{\label{Fig:Halpha_fit} Determination of the effective temperature
    of {\he} from two independent line-profile analyses of H$\alpha$. We 
    estimate the fitting error to be $\sim 70$\,K. For better visibility, we 
    plot line profiles with $\teffm \pm 150$\,K, respectively (dashed lines), 
    corresponding approximately to the estimated \emph{total} error of {\tefft}. 
    See the text for further details.}
\end{figure*}

The effective temperature ({\tefft}) of {\he} was independently derived by two
of us (A.J.K. and P.S.B.) by profile analysis of H$\alpha$ (see Figure
\ref{Fig:Halpha_fit}), yielding $\teffm = 5180\pm 150$\,K and $\teffm = 5140\pm
150$\,K, respectively. The techniques used differ, most importantly, in (a) the
methods used for continuum rectification, (b) the fitting method, (c) the
broadening theory employed -- computations of Stark-broadening by \citep[][
hereafter VCS]{VCS} and self-broadening by \citet{Ali/Griem:1966} versus
\citet{Stehle/Hutcheon:1999} and \citet{Barklemetal:2000} -- , and (d) the model
atmospheres that were used, i.e., a MAFAGS model
\citep{Gehren:1975a,Gehren:1975b}, and a MARCS model, respectively.  In both
cases, $\log g = 2.2$ was assumed, based on the results from section
\ref{Sect:logg}. The effective temperatures derived by A.J.K. and P.S.B. for
{\cd} are $\teffm = 4800\pm 200$\,K and $\teffm = 4710\pm 200$\,K, respectively,
assuming $\log g=1.8$ (NRB01). As already mentioned, the red spectra of {\cd}
suffer from fringing, therefore more accurate determinations were not possible.



We also derived effective temperatures for {\he} from broad- and
intermediate-band photometry. We adopt a reddening of $E(B-V)=0.013$, which was
deduced from the maps of \citet{Schlegeletal:1998}. We note that the maps of
\citet{Burstein/Heiles:1982} yield $E(B-V) = 0.00$. We employed the
color-temperature relations of \citet{Houdasheltetal:2000}, as well as the
empirical relations of \citet{Alonsoetal:1999b,Alonsoetal:2001}, to convert our
de-reddened broad-band visual and infrared colors to estimates of effective
temperatures.  For the Str\"omgren photometry, we used the calibrations of
\citet{Clem:1998}.  The results are summarized in Table
\ref{Tab:TeffDerivation}, together with our results from the H$\alpha$ profile
analysis.

We note that the $J-K$ and $H-K$ colors we observe for {\he} are unusually blue
compared to the observed loci of giant stars and to the theoretical colors of
\citet{Houdasheltetal:2000}. Since the lowest metallicity models included in the
latter have $\mbox{[Fe/H]}=-3.0$, the computations might actually not be
applicable to a star as low in metallicity as {\he}. A change in the observed
$K$ magnitude in {\he} by between $-0.05$ and $-0.10$ magnitudes is required to
make the $V-K$, $J-K$ and $H-K$ consistent with the $V-I$ and $b-y$ colors and
with the temperature derived from H$\alpha$. This effect is not seen in {\cd}
(and also not in other extremely metal-poor stars with $JHK$ measurements we are
aware of). It will be discussed in more detail in a future paper.


Since the $B$ magnitude of {\he} is affected by its strong C overabundance,
and the wavelength range covered by the $K$ band appears to suffer from a flux
deficiency, we neglect the effective temperatures derived from $B-V$ and $V-K$
colors in the determination of the {\tefft}. We adopt the {\tefft} scale
of \citet{Alonsoetal:1999b,Alonsoetal:2001} for our analysis, i.e., $\teffm =
5100\pm 150$\,K, as derived from $b-y$. We emphasize that the chosen {\tefft}
possibly suffers from systematic errors on the order of $\sim 100$\,K.  For
the reader's convenience we provide in the third column of Table
\ref{Tab:AbundanceChanges} the changes of the abundances of all elements we
analyse in {\he} resulting from a change in effective temperature of
$150$\,K.

\subsection{Iron Abundance and Microturbulence}\label{Sect:FeAbundance}


A microturbulent velocity of $v_{\mbox{\scriptsize micr}}=2.2$\,km\,s$^{-1}$ was
determined for {\he} by forcing the abundances of the 25 \ion{Fe}{1} lines we
detect to have no trend with line strength (see Figure \ref{Fig:FeH}).  Since
the number of lines we used is relatively small, and the line-to-line abundance
scatter is considerable (the standard deviation of the sample is $0.15$\,dex), the
accuracy of our determination of $v_{\mbox{\scriptsize micr}}$ is limited. From
the change of the slope of the abundance trend with $W_{\lambda}$ when
changing $v_{\mbox{\scriptsize micr}}$, we estimate that the accuracy of its
determination is $\sim 0.5$\,km\,s$^{-1}$. The average abundance we measure is
$\log \epsilon (\mbox{Fe})=2.06$\,dex, yielding $\mbox{[Fe/H]}=-5.39$, if a
solar iron abundance of $7.45$\,dex is adopted \citep{Asplundetal:2000b}.


We also carried out a differential analysis to {\cd}. From Table
\ref{Tab:TeffDerivation} one can see that the colors that are useful for {\he}
(i.e., neglecting $B-V$ and $V-K$) consistently yield that {\cd} is $\sim
300$\,K cooler than {\he}. The Balmer line-profile analyses yield $\sim 100$\,K
higher temperature differences, on the order of $\Delta\teffm = 400$\,K, which
is in agreement with $\Delta\teffm = 300$\,K to within the measurement
uncertainties. We hence adopt $\teffm=4800$\,K for the differential
analysis. Using this temperature, we determine $\log g=1.8$ and
$v_{\mbox{\scriptsize micr}}=3.4$\,km\,s$^{-1}$ for {\cd}. The differential
analysis of the \ion{Fe}{1} abundance indicates that {\he} is $1.5$\,dex more
iron-poor than {\cd}, which is in good agreement with our LTE value for the
iron abundance of {\he}.

We noted the presence of a trend of $\log\epsilon (\mbox{Fe})$ with $\chi$ (see
Figure \ref{Fig:FeH}). Such a trend is commonly observed in extremely metal-poor
giants \cite[see e.g. NRB01;][]{KeckpaperII}, and the reason for it is unclear.
An investigation of this effect is beyond the scope of this paper. We just note
that it was not possible to remove the trend by changing $v_{\mbox{\scriptsize
micr}}$. Changing the effective temperature has an influence on the trend, but
$\teffm \sim 4700$\,K would have to be adopted for {\he} in order to fully
remove the trend, which is inconsistent with {\tefft} derived from either Balmer
line wings or broad-band photometry (see Table \ref{Tab:TeffDerivation}).

Since there are no Fe~II lines visible in our spectrum of {\he}, we have to
rely on Fe~I lines to derive [Fe/H]. Fe\,{\sc I} (and also Ca\,{\sc I}) has
been suspected to show significant departures from LTE for a long time
\citep{Athay/Lites:1972,Steenbock:1985}. However, recent calculations
\citep{Grattonetal:1999,Thevenin/Idiart:1999} disagree on the validity of LTE
for Fe\,{\sc I} in metal-poor stars.  This discrepancy can be traced back to
the input physics the two groups employ: the different degree of model atom
completeness, different approximations for photoionization coupled with the
treatment of UV fluxes, and the inclusion/neglect of inelastic collisions with
hydrogen atoms.

In a recent series of papers
\citep{Gehrenetal:2001a,Gehrenetal:2001b,Kornetal:2003}, a non-LTE model for
Fe\,{\sc I/II} was presented which addresses all the issues mentioned above. The
model atom was compiled from the recent literature \citep{Naveetal:1994} and
thus reflects the current state of knowledge of the iron term system.
Photoionization is treated by implementing the quantum-mechanical computations
of \citet{Bautista:1997}, giving cross-sections which are systematically larger
than those of previously assumed simple approximations, typically by factors of
100. Line blocking is considered from both continuous and discrete opacity
sources, which allows for a realistic treatment of the UV fluxes as a function
of metallicity. The efficiency of hydrogen collisions is carefully calibrated
using metal-poor stars whose gravities can be inferred from {\sc Hipparcos}
astrometry \citep{Kornetal:2003}. Non-LTE effects for Fe\,{\sc I} turn out to be
intermediate between the ones advocated by the two groups mentioned above, and
amount to $+0.11$\,dex in the case of {\he}. We therefore adopt
$\mbox{[Fe/H]}=-5.3$ for our best estimate of the iron abundance of {\he}.

\subsection{Surface Gravity}\label{Sect:logg}

We use a variety of methods and their combinations to constrain the gravity of
{\he}, including ionization equilibria, relative strengths of Balmer line wings,
and by employing evolutionary tracks. It is a challenging task to derive
spectroscopic gravities for metal-poor stars, even at less extreme metal
deficiencies than that of {\he}, because at $\mbox{[Fe/H]}\lesssim-2$,
individual spectral features in the optical regime (like \ion{Mg}{1} $\lambda
5183$\,{\AA}, \ion{Ca}{1} $\lambda 4226$\,{\AA} or \ion{Fe}{1} $\lambda
4383$\,{\AA}) cease to be strong enough to show a dependence on gravity via
pressure broadening \citep[see][]{Fuhrmann:1998a}.

An independent estimate of the gravity can be derived on the basis of the
ionization equilibrium of e.g. Fe\,{\sc I/II}. However, as already mentioned, no
\ion{Fe}{2} lines are detected in our UVES spectra of {\he}, hence only a lower
limit to the gravity can be derived from these features. Non-LTE corrections for
the \ion{Fe}{1} abundance have been taken into account when deriving these lower
limits. From the non-detection of Fe\,{\sc II} 5169\,{\AA} and 5018\,{\AA} (both
multiplet 42), this limit is determined to be $\log g = 0.0$, if a $3\,\sigma$
detection limit according to Equation (\ref{Eq:sigWlambda}) is used, and $\log g
= 1.3$ in case of a $1\,\sigma$ limit.  A stronger constraint follows from a
pixel-by-pixel comparison of model spectra with the observed spectrum in the
wavelength regions occupied by the two \ion{Fe}{2} lines (see Figure
\ref{Fig:FeIIsynth}). Gravities below 2.0\,dex are excluded, considering that
the synthesized \ion{Fe}{2} lines for this gravity are outside of the
$1\,\sigma$ range of \emph{all} 6 pixels of the observed spectrum covering both
of these lines. However, confirmation of this result by higher $S/N$ spectra is
needed. The situation could also be remedied by analysing additional Fe\,{\sc
II} lines around 3250\,{\AA}, which are significantly stronger than any of the
lines at longer wavelength.


Main-sequence gravities can be excluded for {\he} by further constraints
obtained from the Balmer line-profile analysis. Although the wings of H$\alpha$
are mainly sensitive to effective temperature, there is also a slight gravity
dependence.  Inspection of a 12\,Gyr, $Z=10^{-5}$ isochrone \citep{Yietal:2001}
suggests that a star of $\teffm = 5100$\,K in a pre-helium core flash
evolutionary state either has $\log g=2.2$ (i.e., if it is on the red giant
branch) or $\log g=4.8$ (i.e., if it is on the main sequence). Assuming the
latter in a profile analysis of H$\alpha$, and all other input as in \S
\ref{Sect:Teff} for the analysis of P.S.B., yields $\teffm = 4700$\,K. Such a
low temperature is inconsistent with broad-band photometry (see Table
\ref{Tab:TeffDerivation}).

We can also rule out dwarf-like gravities by applying the constraint that
effective temperatures derived from different Balmer lines should be in
agreement. For cool stars like {\he}, the sensitivity of the higher Balmer line
wing depths to gravity is in the opposite sense to that found for lower lines in
the series, such as H$\alpha$.  This is due to the fact that self-broadening is
more important than Stark broadening for the lowest lines in the series, while
Stark broadening completely dominates for higher members of the series.  For a
purely self-broadened H line, the line opacity in the wings is proportional to
the number density of H-atoms, and this can be shown to lead to a sensitivity
such that line strength increases with gravity. For a purely Stark broadened
line, while the wing opacity due to static perturbers scales with electron
density, the wing opacity due to electrons in the impact regime does not, and
this can be shown to lead to line strength decreasing with increasing gravity.
The sensitivity of the line strength to gravity is in fact reasonably strong for
approximately $\log g > 2$.  Using this behavior, simultaneous fitting of
H$\alpha$ and one or more of the higher Balmer lines (we found H$\gamma$ and
H$_{10}$ to be most useful in {\he}), can constrain {\tefft} and $\log
g$. $T_\mathrm{eff}\approx4700$ and $\log g=4.8$ is completely inconsistent with
the observed H$\gamma$ and H$_{10}$ spectra, yielding synthetic profiles that
are much too weak, while the profiles obtained from the combination of
parameters $T_\mathrm{eff}\sim 5100$ and $\log g=2.2$ are consistent with the
observations (see Figure \ref{Fig:HalphaH10}).

\begin{figure*}[htbp]
  \epsfig{file=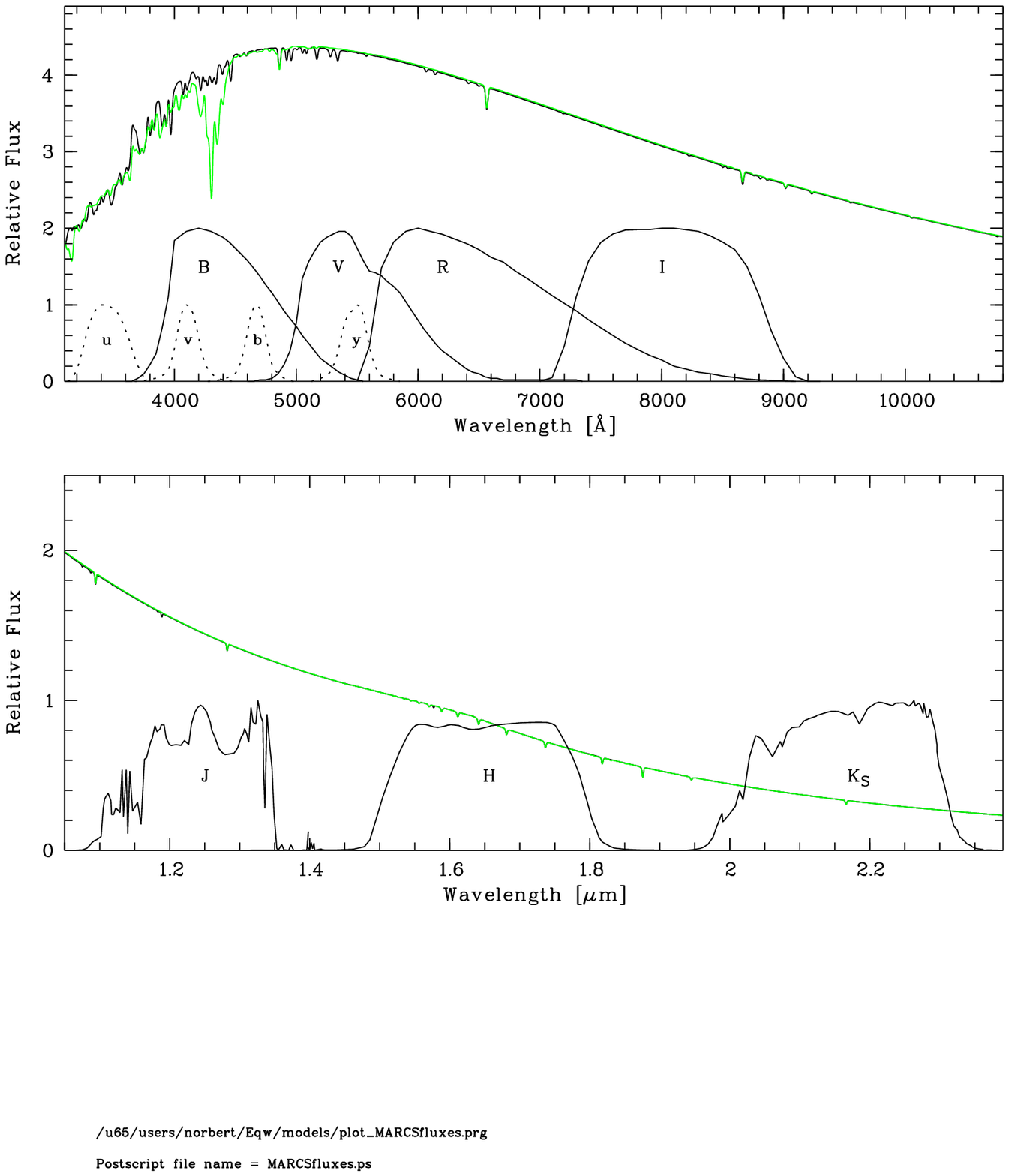, clip=, width=14cm,
    bbllx=55, bblly=212, bburx=513, bbury=628}
  \centering
  \caption{\label{Fig:MARCSfluxes} Relative fluxes of MARCS models, smoothed
    with a 20\,{\AA} FWHM Gaussian, together with broad- and intermediate-band filter
    transmission functions. Black line: A model spectrum with
    $\mbox{[Fe/H]}=-3.0$, a scaled solar composition, and $\alpha$-enhancement
    of 0.5\,dex. Grey line: A model spectrum with $\mbox{[Fe/H]}=-5.4$, tailored
    for the abundance pattern of {\he}. In particular, the large carbon
    enhancement is taken into account in that model, as can be seen from the
    strong G band of CH at $\sim 4300$\,{\AA}.  While the magnitudes in the
    $u$, $v$, and $B$ filters are influenced by the large C enhancement of
    {\he}, the redder magnitudes are not. }
\end{figure*}

\begin{figure*}[htbp]
  \epsfig{file=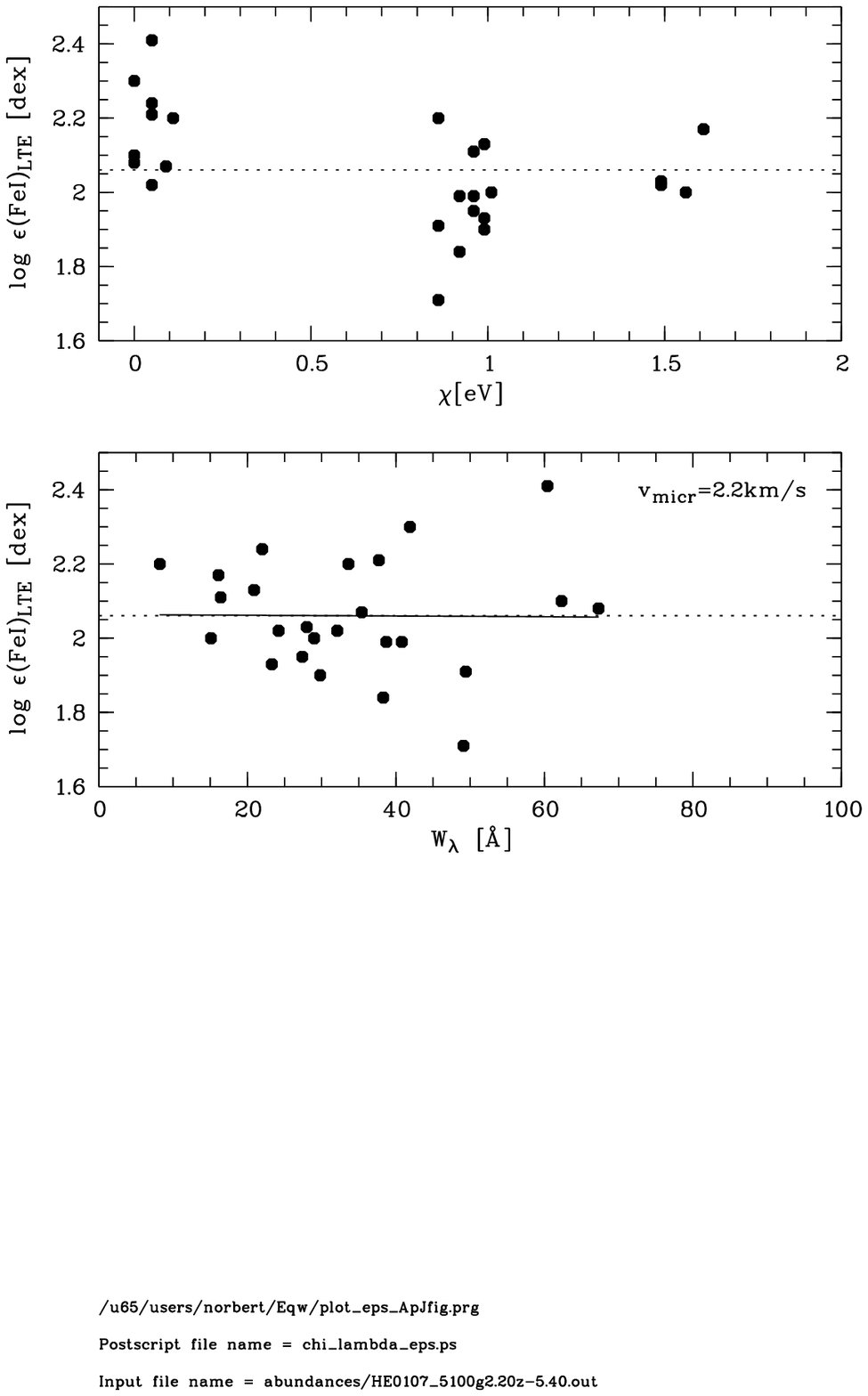, clip=, height=14cm,
    bbllx=75, bblly=284, bburx=408, bbury=613}
  \centering
  \caption{\label{Fig:FeH} LTE iron abundance, $\log\epsilon (\mbox{Fe})$, of
    {\he} as a function of excitation potential $\chi$ (upper panel) and
    equivalent width $W_{\lambda}$ (lower panel), for the adopted stellar
    parameters $\teffm = 5100$\,K, $\log g=2.2$, and $v_{\mbox{\scriptsize
        micr}}=2.2$\,km\,s$^{-1}$. Note the presence of a trend of the
    abundance with $\chi$. This is commonly observed in extremely metal-poor
    giants (see e.g., NRB01). See text for further discussion. }
\end{figure*}

\begin{figure*}[htbp]
  \epsfig{file=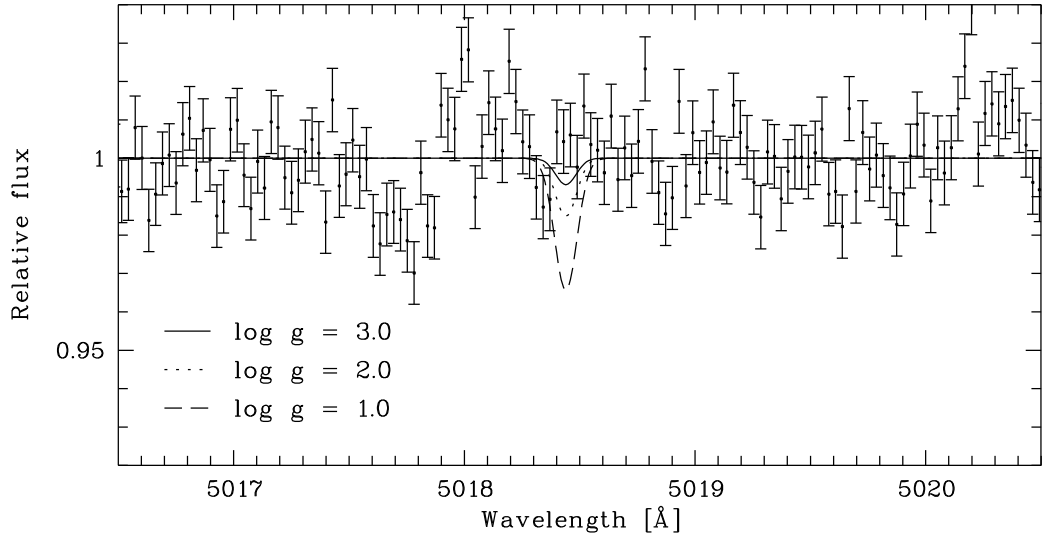, clip=, width=14cm,
    bbllx=70, bblly=425, bburx=458, bbury=629}
  \centering
  \caption{\label{Fig:FeIIsynth} Spectrum synthesis of the wavelength regions
    occupied by \ion{Fe}{2} $\lambda 5018.440$\,{\AA}.
    }
\end{figure*}

\begin{figure*}[htbp]
  \epsfig{file=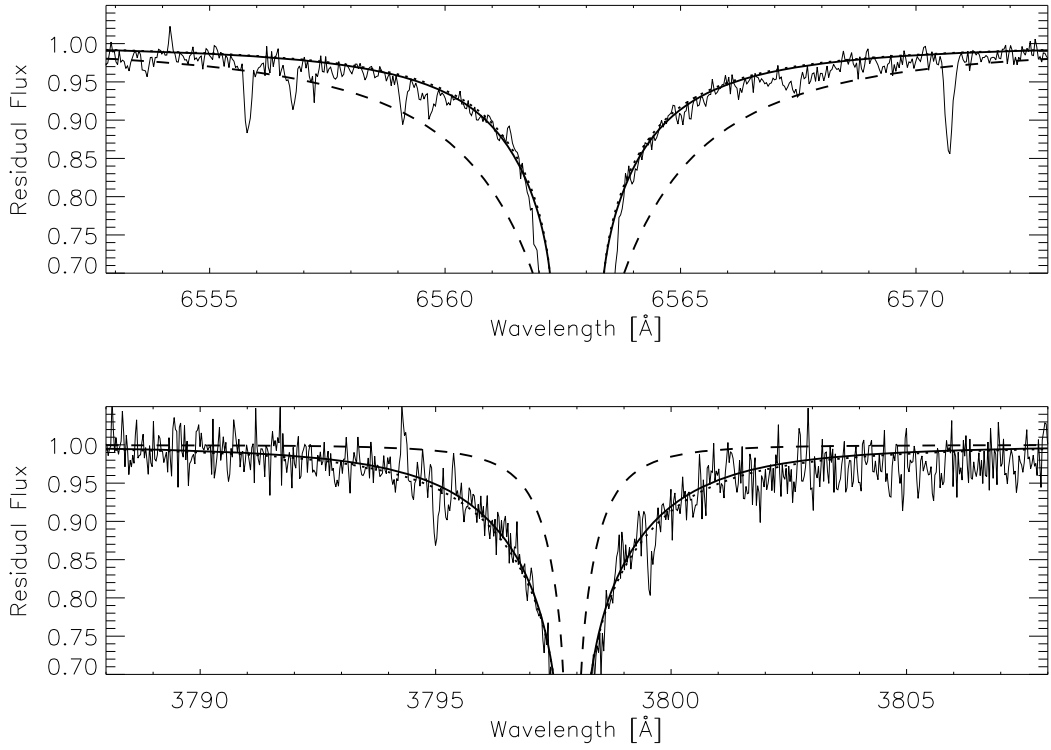, clip=, width=14cm,
    bbllx=70, bblly=365, bburx=542, bbury=704}
  \centering
  \caption{\label{Fig:HalphaH10} Comparison of observed (thin full line) and
    model spectra for H$\alpha$ (top panel) and H$_{10}$ for fixed
    $\teffm=5140$\,K and $\log g=1.8$ (dotted line), $\log g=2.2$ (thick full
    line) and $\log g=4.8$ (dashed line).  We can easily rule out
    the possibility that {\he} is a main sequence star. A gravity typical of a
    giant, $\log g=2.2$, is clearly consistent with the observations.
    }
\end{figure*}

Finally, although we note that our measured value of the gravity sensitive index
$c_1=0.09$ yields a dwarf-like gravity, when we employ the Str\"omgren color
calibrations of \citet{Clem:1998}, this index is disturbed by the strong CH
lines that are present in the wavelength ranges covered by the $u$ and $v$
filters (see Figure \ref{Fig:MARCSfluxes}). In particular, the flux in the $v$
filter is suppressed by the G band of CH, which leads to higher values for the
$b$ magnitude, and since $c_1=(u-v)-(v-b)$, a low $c_1$ color index results,
yielding a high gravity. Hence a calibration of $c_1$ that takes into account
the peculiar abundance pattern of {\he} needs to be established before any $c_1$
measurements are useful for gravity determination. Such calibration efforts are
in progress.

We have combined our constraints from the above mentioned indicators (with the
exception of $c_1$) with pre-helium flash stellar evolutionary tracks. This
yields $\log g = 2.2\pm 0.3$, with the uncertainty arising mainly from the
uncertainty in {\tefft}, and we shall adopt this value for the present analysis.

\section{ABUNDANCE ANALYSIS}\label{Sect:AbundAnalysis}

\subsection{Solar Abundances}\label{Sect:SolarAbundances}

For the computation of elemental abundances relative to the Sun, as listed in
Table \ref{Tab:Abundances} below, we mostly adopted the solar abundances of
\citet[][ hereafter GS98]{Grevesse/Sauval:1998}, with a few exceptions
described in the following.

\citet[][ henceforth H01]{Holweger:2001} lists improved solar abundances of C,
N, O, Ne, Mg, Si, and Fe, derived by taking into account non-LTE effects and
granulation. The latter is realized by a 2D approach; for the vertical
temperature structure, the solar model of \citet{Holweger/Mueller:1974} is
employed. Of the H01 solar abundances relevant for our analysis, we adopt
those of N and Mg, which differ by the values listed in GS98 by $+0.01$ and
$-0.04$\,dex, respectively, in the sense H01$-$GS98.

For C, Si, and Fe, solar abundances derived by 3D hydrodynamical simulations are
available. Comparison with observations have shown that the simulations of
Asplund~et~al. appear to be realistic, in that they reproduce the shapes,
shifts, and asymmetries of solar Fe lines very well
\citep{Asplundetal:2000a}. \citet{Asplundetal:2000b} derive $\log\epsilon
(\mbox{\ion{Fe}{1}})_{\odot}=7.44\pm 0.05$ and $\log\epsilon
(\mbox{\ion{Fe}{2}})_{\odot}=7.45\pm 0.10$ by means of such simulations, and we
adopt their \ion{Fe}{2} value.  Note that this value agrees very well with the
meteoritic iron abundance, if a downward correction of the solar photospheric Si
abundance of $0.04$\,dex with respect to GS98 is taken into account
\citep{Asplund:2000}. We also adopt the solar Si abundance of Asplund, i.e.,
$\log\epsilon (\mbox{Si})_{\odot}=7.51\pm 0.04$. Finally, again using 3D models,
\citet{AllendePrietoetal:2002} obtain $\log\epsilon (\mbox{C})_{\odot}=8.39\pm
0.04$ from the \ion{C}{1} 8727\,{\AA} line, for which they show non-LTE effects
to be negligible. We adopt this value, which is $0.13$\,dex lower than the value
of GS98, and $0.2$\,dex lower than the value of H01.

\subsection{Line Identification}

Since numerous lines of CH are present in the spectrum of {\he}, line
identification is not as easy as for other extremely metal-poor stars. We
started the identification process with line lists used in recent
high-resolution studies of extremely metal-poor stars
\citep{McWilliametal:1995b,Norrisetal:1996,Norrisetal:2001,Depagneetal:2000,KeckpaperII},
complemented by \citet{Bessell/Norris:1984}, and the solar atlas
\citep{Mooreetal:1966}. Additional identifications were made using the
Vienna Atomic Line Database\footnote{\texttt{http://www.astro.uu.se/$\sim$vald/}}
\cite[VALD;][]{VALD2a,VALD2b}.

Apart from checking the consistency of the abundances of a given element derived
from different lines, we carried out careful checks for all species for which we
detected only a few, or even only single lines. The result is that, apart from
hydrogen lines, and numerous molecular lines of CH, CN and C$_2$, we identified
41 lines of 6 elements (Na, Mg, Ca, Ti, Fe, and Ni), and determined upper limits
for lines of 12 additional elements. The results are summarized in Table
\ref{Tab:Linelist}.


 
\subsection{Measurement of Equivalent Widths}\label{Sect:EqwMeasurement}

Equivalent widths were measured from the rebinned UVES spectra with a
semi-automatic procedure. In a 4\,{\AA} wide region centered on the line to be
measured, a simultaneous $\chi^2$ fit of a straight line continuum and a
Gaussian is performed. While the use of a Gaussian leads to systematic
underestimation of the equivalent widths of strong lines due to the presence of
damping wings, which cannot adequately be reproduced by a Gaussian profile, this
is not an issue in case of {\he}, since the strongest line we measured with this
method has $W_{\lambda}=67$\,m{\AA}. The equivalent width measurements and upper
limits for {\he} we measured are summarized in Table \ref{Tab:Linelist}.


\subsection{The Model Atmospheres}\label{Sect:Modelatm}

The model atmospheres used in the analysis were calculated using the present
version of the MARCS program. This program has been developed through a number
of different versions
\citep{Gustafssonetal:1975,Plezetal:1992,Edvardssonetal:1993,Jorgensenetal:1992,Asplundetal:1997}.
A more detailed description is presently being prepared by Gustafsson et al. The
MARCS program calculates predicted model atmospheres for late-type stars. It is
based on the assumptions of stratification in plane-parallel or
spherically-symmetric layers, hydrostatic equilibrium, mixing-length convection,
and LTE.  The continuous and line absorption from atoms and molecules is
considered in full detail by opacity sampling with $2\times 10^4$ to $2\times
10^5$ wavelength points. 

The parameters of the model used in the final iteration of the present
analysis are given in Table \ref{Tab:StellarParameters}, with the exception of
[Fe/H], for which we use $\mbox{[Fe/H]}=-5.4$ for consistency reasons. The
model takes into account the strong overabundances of C and N in {\he}, i.e.,
$\log\epsilon (\mbox{C})=7.4$\,dex and $\log\epsilon (\mbox{N})=4.7$\,dex, as
determined in a close-to-final iteration, were used. An enhancement of
$\alpha$-elements, including O, by $+0.5$\,dex was assumed, and the MLT
parameters $\alpha=l/H_p$ and $y$ were set to $1.5$ and $0.076$, respectively.

\begin{figure*}[htbp]
  \begin{center}
    \begin{turn}{90}
    \epsfig{file=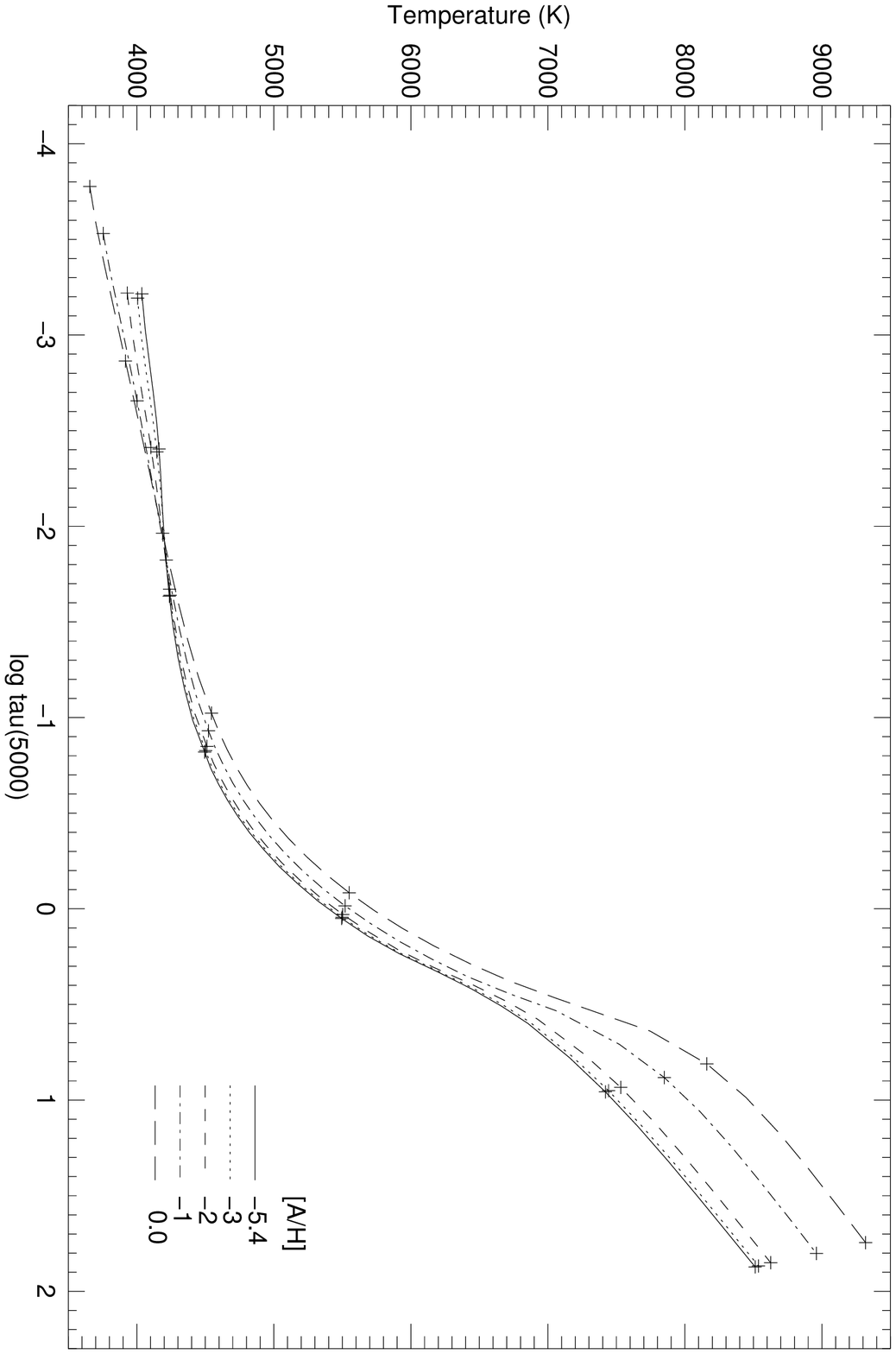, clip=, height=14cm,
      bblly=72, bbllx=78, bbury=736, bburx=508}
    \end{turn}
    \begin{turn}{90}
    \epsfig{file=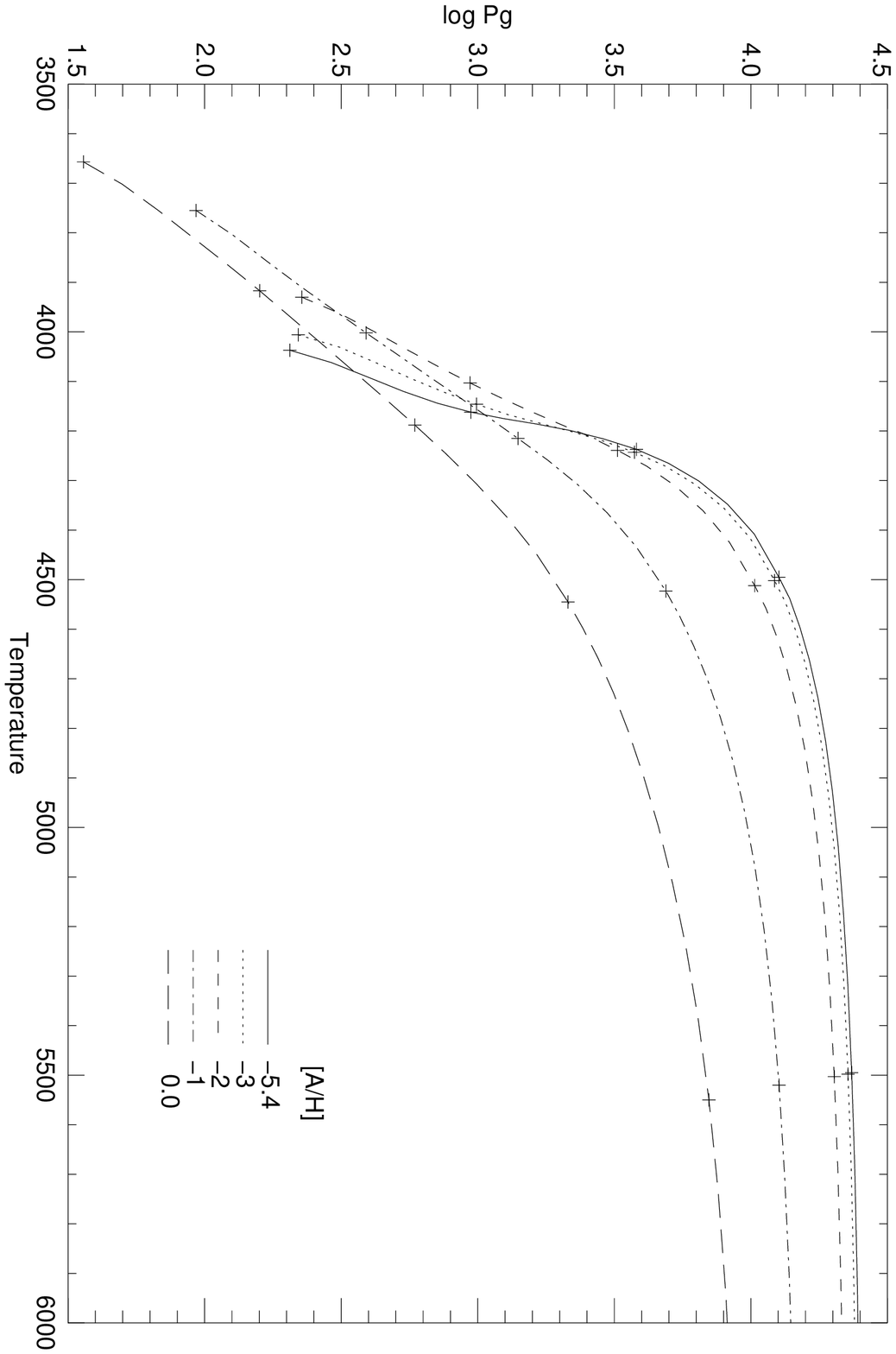, clip=, height=14cm,
      bblly=72, bbllx=78, bbury=736, bburx=508}
    \end{turn}
  \end{center}
  \centering
  \caption{\label{Fig:MARCSmodels} MARCS model atmospheres calculated for
    $\teffm = 5\,100$\,K, $\log g = 2.2$, and $\mbox{[Fe/H]}$ ranging from
    $-5.4$ to solar. The carbon and nitrogen abundances were the C/H and N/H
    abundances determined for {\he} if they were smaller than the
    corresponding solar abundances, scaled to the overall metallicity of the
    model. If not, the latter were adopted. Crosses mark the points in the
    models where $\log \tau_{\mbox{\scriptsize Ross}} = -4$, $-3$, $-2$, $-1$,
    $0$, and $1$, respectively. Upper panel: Temperature versus logarithmic
    optical depth at $\lambda = 5\,000$\,{\AA}; Lower panel: Logarithmic gas
    pressure versus temperature.
 }
\end{figure*}

The model structure is given in Table \ref{Tab:MARCSmodel} and compared with a
set of more metal-rich models ([A/H] denoting the over-all metal abundance) with
the same {\tefft} and $\log g$ in Figure \ref{Fig:MARCSmodels}. It is seen that
the pressures at a given temperature and optical depth increase as the
metallicity decreases, as expected. Note that the models approach an asymptotic
structure when [A/H] approaches $-3$. The reason for this is that
line-blanketing effects vanish as the metallicity decreases. The only
significant opacities in the most metal-poor models are the H$^-$ absorption,
Rayleigh-scattering, absorption by \ion{H}{1}, and the H$_2^+$ absorption in the
blue and near ultraviolet, where it amounts to $< 20$\,\% of the H$^{-}$
absorption. In the line-forming regions (around $\tau_{\mbox{\scriptsize
Ross}}=0.01$), \ion{H}{1} scattering becomes stronger than the H$^-$ absorption
shortwards of about 5\,000\,{\AA}; at longer wavelengths and at greater depths,
H$^-$ dominates until \ion{H}{1} absorption takes over below
$\tau_{\mbox{\scriptsize Ross}}=5$. (In the far UV, \ion{H}{1} absorption is
also important, as is C~I absorption, but at the low effective temperature of
{\he}, this has only minor consequences for the structure of the model.) The
line blanketing effects, mainly due to \ion{H}{1} and CH absorption, are small.

One might suspect that metals could be of significance as electron donors, but
this is not so, since throughout the model hydrogen is the totally dominant
electron donor, contributing more than 99\,\% of the electrons.  In fact, we
find from the sequence of models displayed in Figure \ref{Fig:MARCSmodels} that
the metals do not provide electrons comparable to the contribution from hydrogen
in the line-forming regions for $\mbox{[A/H]} \lesssim -2$.  Similarly, only for
$\mbox{[A/H]} \gtrsim -2$ does the H$^{-}$ opacity become more important than
\ion{H}{1} scattering in the blue and violet for the line-forming regions.

The gas pressure is entirely dominated by the contributions from atomic hydrogen
and helium. At the surface gravity adopted for {\he}, H$_2$ molecules contribute
less than $0.1$\,\% of the total pressure. Among other molecules, CH is the most
important pressure contributor for the chemical composition chosen.  Note that
CO is less significant, in spite of the low temperature of the gas, which for
more metal-rich gas would mean that a great fraction of available oxygen would
form CO molecules, as a result of the high dissociation energy of this
molecule. This is not the case at these low metallicities, where most of the
oxygen, and also most of the carbon, is still in atomic form. The relatively
small number of CO molecules reflects the fact that the partial pressure of CO
is proportional to the product of the partial pressures of C and O atoms,
respectively.

The model atmosphere adopted for {\he} is unstable against convection for
$\tau_{\mbox{\scriptsize Ross}} > 0.4$, and convection carries more than
50\,\% of the total flux for $\tau_{\mbox{\scriptsize Ross}} > 6$, according
to mixing-length theory \citep{BohmVitense:1958}. The fact that
convection reaches visible layers is basically due to the transparency of the
metal-poor gas; a consequence of the absence of metals as electron donors,
which leads to less H$^-$ formation at a given pressure and temperature than
in corresponding more metal-rich models. The fact that the model is
convectively unstable in visible layers makes the present analysis uncertain;
mixing-length theory is known to be inadequate even for describing a mean
structure of the atmosphere. This will be discussed further below.

\subsection{Elemental Abundances of {\he}}

\begin{figure*}[htbp]
  \epsfig{file=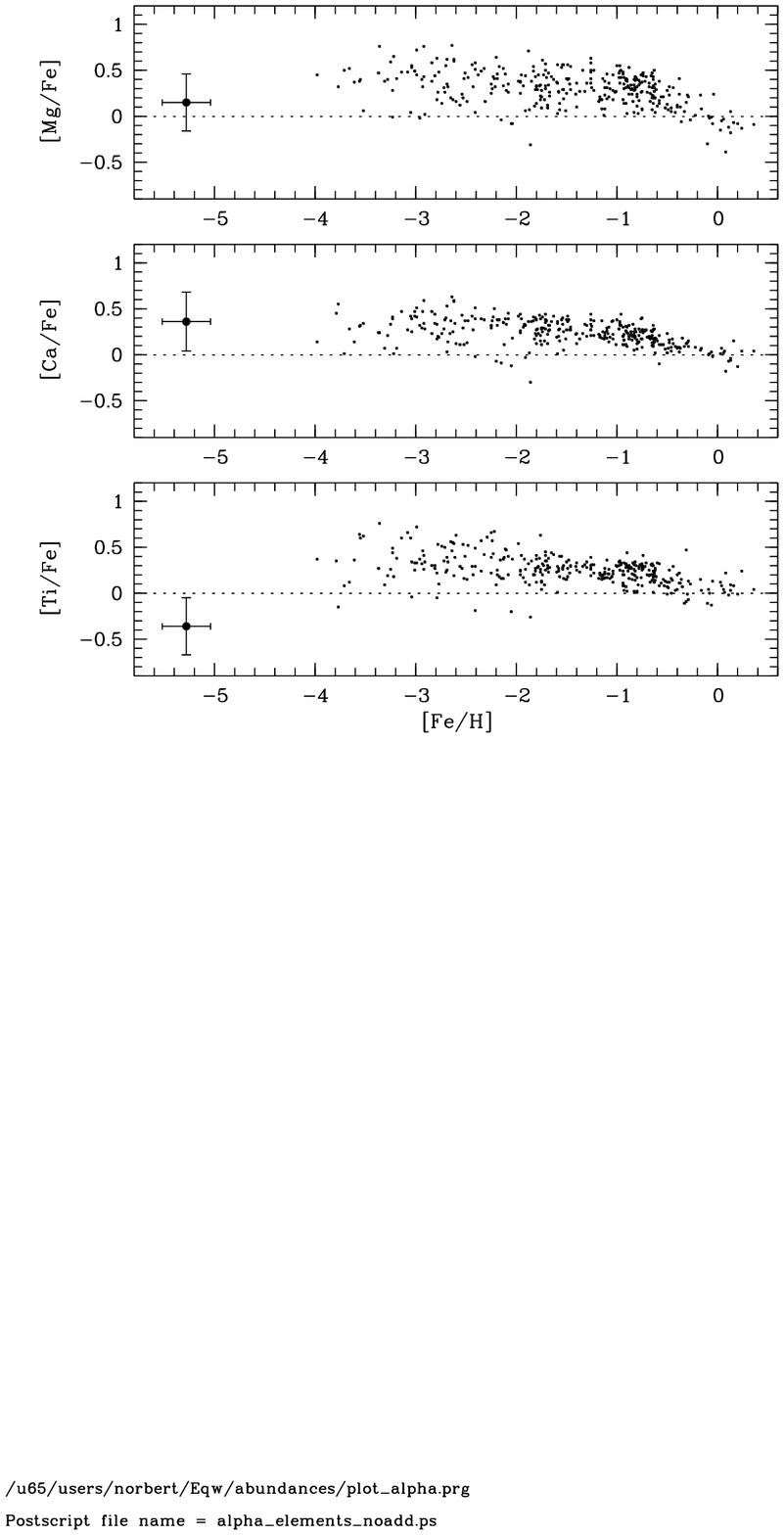, clip=, width=14cm,
    bbllx=125, bblly=415, bburx=456, bbury=740}
  \centering
  \caption{\label{Fig:AlphaElementRatios} Abundance ratios of the
    $\alpha$-elements as a function of [Fe/H] for extremely metal-poor stars,
    and {\he} (filled circle). The error bars for {\he} include the total errors
    listed in Table \ref{Tab:AbundanceChanges} quadratically added to an assumed
    $1\,\sigma$ error of $0.1$\,dex from uncertainties in atomic data. The
    abundances for the stars with $\mbox{[Fe/H]}>-4.0$ in this figure as well as
    in Figure \ref{Fig:NiFe} were taken from NRB01, and were
    kindly provided in digital form by the authors.}
\end{figure*}

\begin{figure*}[htbp]
  \epsfig{file=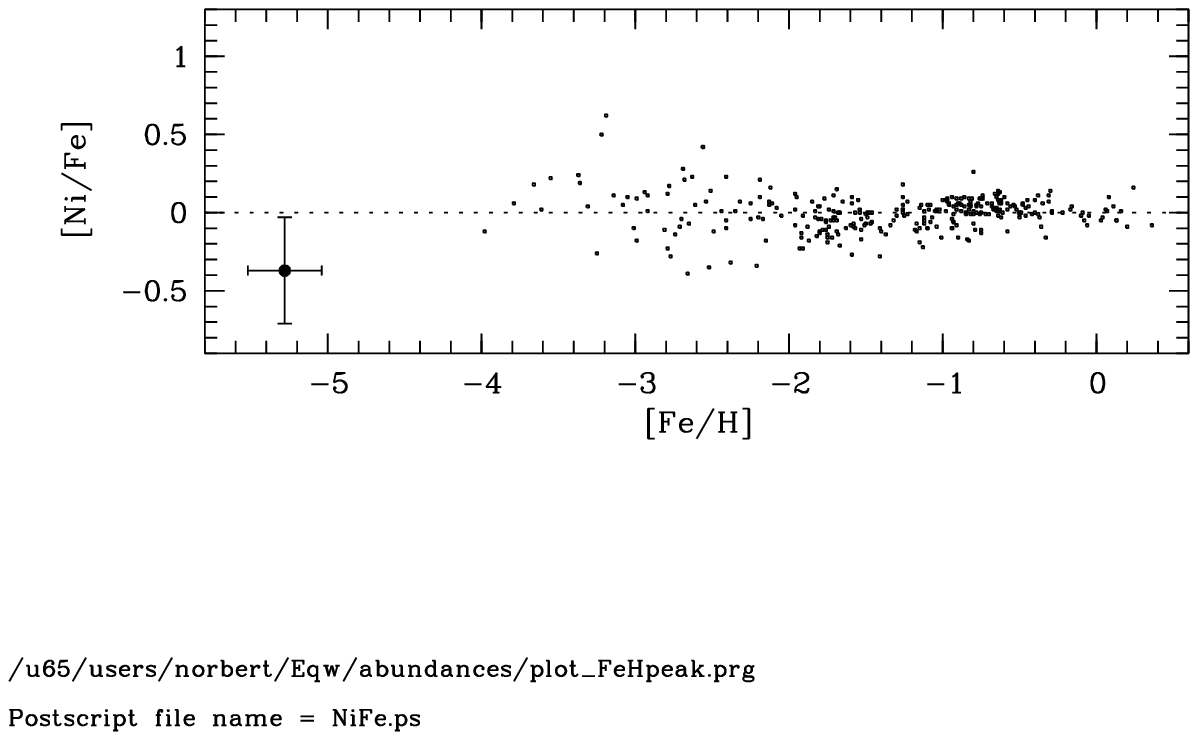, clip=, width=14cm,
    bbllx=125, bblly=147, bburx=456, bbury=278}
  \centering
  \caption{\label{Fig:NiFe} [Ni/Fe] as a function of [Fe/H].}
\end{figure*}

The elemental abundances we derive for {\he} are summarized in Table
\ref{Tab:Abundances}, and are shown in comparison with abundances of other
metal-poor stars in Figures \ref{Fig:AlphaElementRatios} and \ref{Fig:NiFe}. In
this section we comment on elements that required special attention in the
analysis.

\subsubsection{Carbon}

\begin{figure*}[htbp]
  \epsfig{file=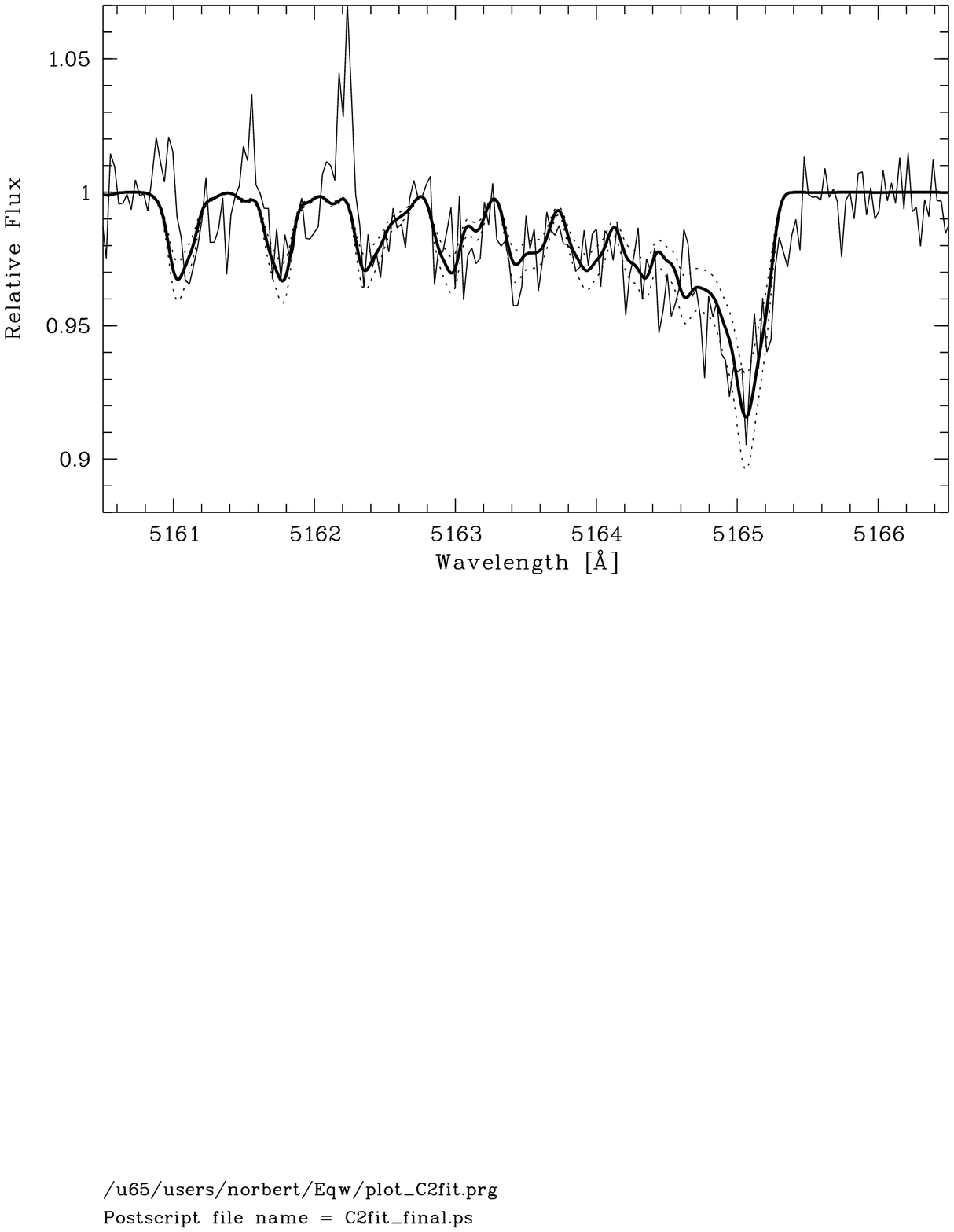, clip=, width=14cm,
    bbllx=46, bblly=420, bburx=528, bbury=713}
  \centering
  \caption{\label{Fig:C2} Spectrum synthesis of the C$_2(\mbox{A}^3\Pi\mbox{--}\mbox{X}^3\Pi)$
    feature at $\sim 5165$\,{\AA} in {\he}. Thin solid line: observed
    spectrum; thick solid line: best fit, $\log\epsilon(\mbox{C})=7.11$;
    dashed lines: $\log\epsilon(\mbox{C})=7.06$ and $\log\epsilon(\mbox{C})=7.16$,
    respectively.
    }
\end{figure*}

We performed a spectrum synthesis of the C$_2\,(0,0)$ Swan band at $\sim
5165$\,{\AA} (see Figure \ref{Fig:C2}), and moderately-strong lines of the
CH~A-X system in the wavelength regions $4210$--$4225$\,{\AA},
$4283.5$--$4290.5$\,{\AA}, and $4361.5$--$4373.5$\,{\AA}. The best fits were
determined by minimizing $\chi^2$ between the synthesized and observed
spectra. In the case of CH, wavelength regions containing lines that are missing
in our CH line list were excluded from the fit. The spectrum synthesis yielded C
abundances discrepant by 0.3\,dex between both indicators, i.e., $\log\epsilon
(\mbox{C})=7.1$\,dex from C$_2$, and $\log\epsilon (\mbox{C})=6.8$\,dex from
CH.

Such a discrepancy has been found before in extremely metal-poor,
carbon-enhanced giants by many previous authors.  For example,
\citet{Hilletal:2000} found a discrepancy of $0.15$\, dex for CS~22948-027 and
CS~29497-034, when they used moderately strong CH lines at $\sim 4000$\,{\AA}.
They obtained a discrepancy of $\sim 0.6$\,dex when they used the strong lines
of the G band at $4300$\,{\AA}.  \citet{Bonifacioetal:1998} have seen the same
effect in CS~22957-027.

Due to the different sensitivity of C$_2$ and CH to temperature and gravity (see
Table \ref{Tab:AbundanceChanges}) the discrepancy in {\he} could be removed if
$\teffm\sim 5500$\,K or $\log g=0.8$\,dex were assumed. However, both values
are clearly excluded (see \S \ref{Sect:StellarParameters}), and furthermore, the
fact that such a discrepancy has been observed in other extremely metal-poor
stars as well suggests that the reason is not due to peculiarities in {\he}, but
among common systematic errors of the abundance analysis methods used, e.g.,
inadequate temperature structures of 1D models. In fact, as discussed in \S
\ref{ModelatmErrors} below, 3D hydrodynamical simulations yield much lower
temperatures in the outer layers of the atmosphere than 1D models. Hence, it is
expected that molecular lines in particular may be affected. Preliminary
calculations of Asplund (2003, private communication) indeed suggest that the 3D
corrections for C$_2$ are on the order of 0.2--0.4\,dex larger than those for
CH, at least for metal-poor dwarfs similar to the Sun. Although these 3D
calculations are not tailored for {\he}, one would expect qualitatively similar
behavior in giants and dwarfs, which could therefore possibly bring the C
abundances derived from both indicators into better agreement in {\he}.


From spectrum synthesis of $^{13}\mbox{CH}$ in the three wavelength regions
mentioned above we determine the carbon isotopic ratio
$^{12}\mbox{C}/^{13}\mbox{C}$. The $^{12}$CH and $^{13}$CH line lists were
supplied by J{\o}rgensen \citep[see][]{Jorgensenetal:1996}. A comparison with
laboratory wavelengths, as well as with the analysis of
\citet{Krupp:1973,Krupp:1974} and a related line list of R.A. Bell (private
communication), and the recent measurements and anlysis of the $^{12}$CH and
$^{13}$CH A-X system by \citet{Zachwieja:1997}, showed that the wavelengths of
the $^{12}$CH lines of the A-X system agreed fairly well, while the $^{13}$CH
lines were misplaced by typically $0.27$\,{\AA} in the 4200\,{\AA} region. We
therefore adopted the wavelengths of the $^{13}$CH lines of Bell's lists, but
note that there are individual departures between those and Zachwieja's tables
of $0.005$--$0.08$\,{\AA}.

\begin{figure*}[htbp]
  \epsfig{file=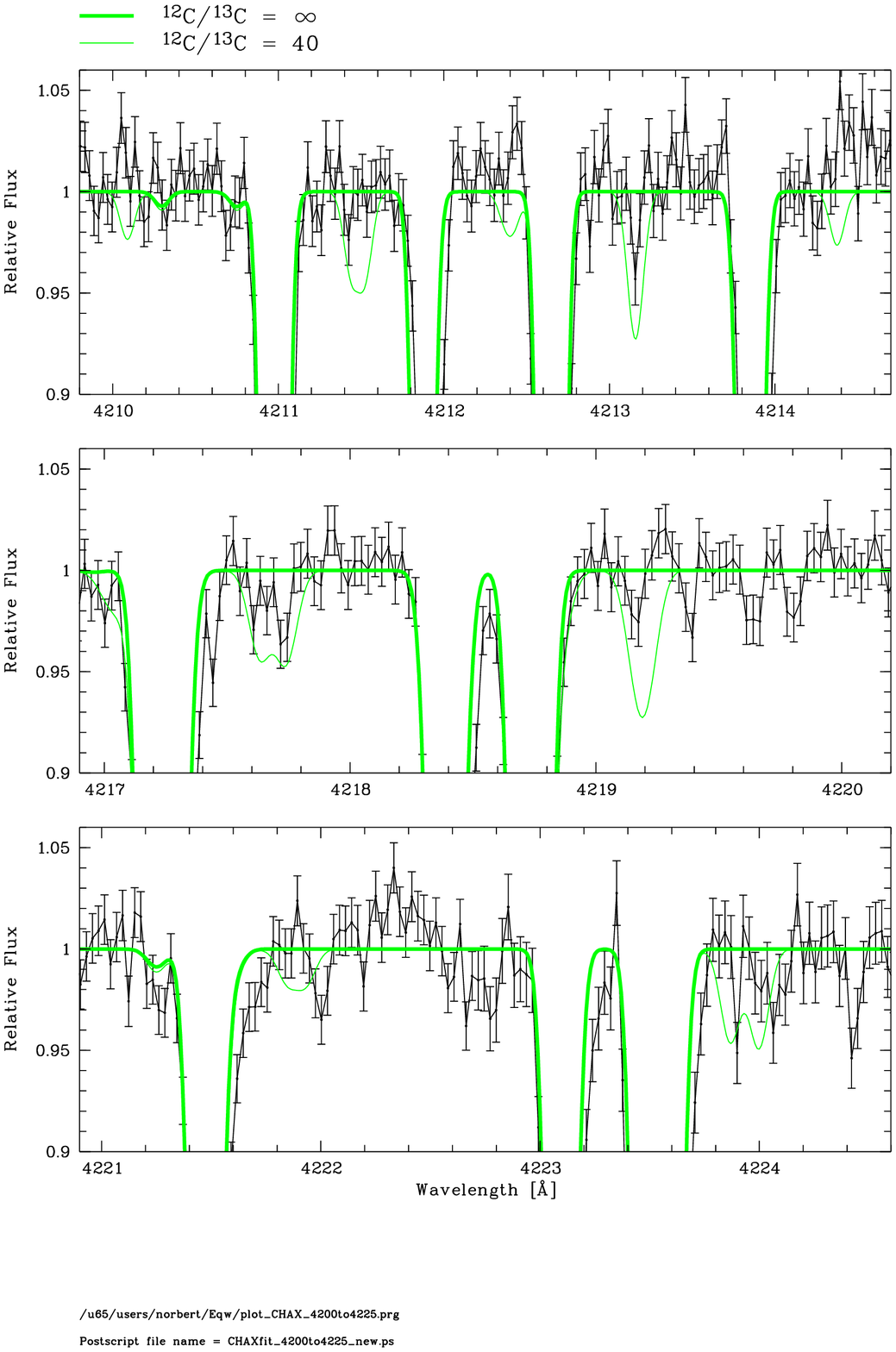, clip=, width=14cm,
    bbllx=55, bblly=113, bburx=527, bbury=713}
  \centering
  \caption{\label{Fig:C12C13fit} Spectrum synthesis of CH-AX lines with
    different carbon isotopic ratios. Thick grey line:
    $^{12}\mbox{C}/^{13}\mbox{C}=\infty$; thin grey line:
    $^{12}\mbox{C}/^{13}\mbox{C}=40$. By means of the displayed spectra (thin
    black line, shown together with $1\,\sigma$ error bars)
    $^{12}\mbox{C}/^{13}\mbox{C}<50$ can be excluded for {\he}. This is
    in concert with additional constraints from other CH lines visible in
    the UVES spectra. For further discussion see text.
    }
\end{figure*}

We have probably detected a $^{13}\mbox{CH}$ feature at $4217.6$\,{\AA}, and a
few other features, e.g. at $4211.4$\,{\AA}, $4212.4$\,{\AA}, and
$4213.2$\,{\AA} also seem to be present (see Figure \ref{Fig:C12C13fit}). This
detection is not definitive, however, in view of the fact that the wavelengths
still must be regarded as uncertain, on the basis of the comparison between
different line lists. If these detections are real,
$^{12}\mbox{C}/^{13}\mbox{C}\sim 60$ would result, a value with considerable
uncertainty also due to the uncertainty in the continuum placement.  However, we
can rule out $^{12}\mbox{C}/^{13}\mbox{C}<50$. Higher quality spectra
are needed to clarify the situation.

\subsubsection{Nitrogen}

\begin{figure*}[htbp]
  \epsfig{file=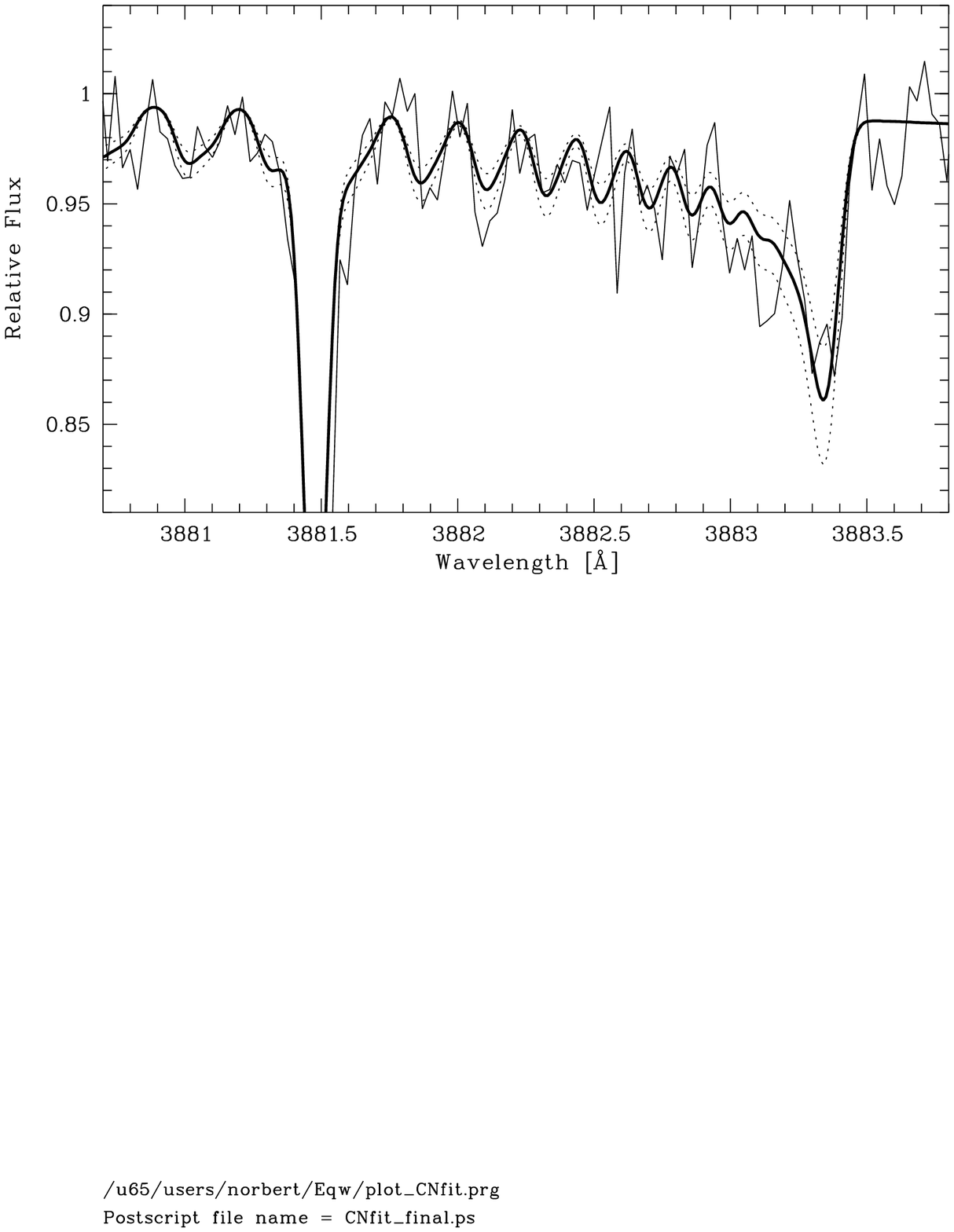, clip=, width=14cm,
    bbllx=46, bblly=420, bburx=528, bbury=713}
  \centering
  \caption{\label{Fig:CN} Spectrum synthesis of the (0,0) band head of the CN
    violet system.  Thin solid line: observed spectrum; thick solid line:
    $\log\epsilon(\mbox{N})=5.22$, which is the best fit for an assumed carbon
    abundance of $\log\epsilon(\mbox{C})=6.81$; dashed lines:
    $\log\epsilon(\mbox{N})=5.12$ and $\log\epsilon(\mbox{N})=5.32$.  }
\end{figure*}

The nitrogen abundance of {\he} was derived by spectrum synthesis of the
violet CN system with the $(0,0)$ band head at 3883\,{\AA} (see Figure
\ref{Fig:CN}). The nitrogen abundance as measured from CN depends linearly on
the assumed C abundance.  Since it was not possible to derive a unique
abundance of C, we used the two different C abundances obtained from C$_2$ and
CH, $\log \epsilon (\mbox{C})=7.1$\,dex and $6.8$\,dex, respectively, in the
spectrum synthesis of CN.




\subsubsection{Calcium}

In order to determine the Ca abundance of {\he} we performed a spectrum
synthesis of the \ion{Ca}{2}~K line. It is very strong and shows damping wings,
and would therefore not be treated adequately with our equivalent width
measurement program. We used the $\log gf$ value from VALD ($\log gf=0.105$),
which is well-confirmed by theory and experiment \citep[see,
e.g.,][]{Theodosiou:1989}. The \ion{Ca}{1} line at $\lambda\,4226.73$\,{\AA} is
unfortunately blended with a CH line at the resolution of our spectra, thus the
Ca abundance derived from this line is of limited reliability.

\subsection{Uncertainties in Abundances}\label{Sect:AbundanceErrors}

A number of different circumstances contribute uncertainties to the abundances
derived. Among these are errors in the measured equivalent widths, in the basic
physical data such as $\log gf$ values or dissociation energies, in the
fundamental stellar parameters adopted, in the model atmospheres, and in the
spectrum calculations. We here comment on these different uncertainties. The
changes of abundances resulting from variations of stellar parameters and
in equivalent width are summarized in Table \ref{Tab:AbundanceChanges}.

\subsubsection{Errors in Measured Equivalent Widths}

From Equation (\ref{Eq:sigWlambda}) and the quality of the UVES spectra it
follows that the $1\,\sigma$ uncertainties in the measurement of equivalent
widths $W_{\lambda}$ of \emph{individual} lines is $\sim 3$\,m{\AA} throughout
most of the wavelength range covered; in the very blue part it continuously
rises from $\sim 5$\,m{\AA} at 3700\,{\AA} to $\sim 20$\,m{\AA} at 3300\,{\AA}.
The precision in abundance determination can clearly be improved if many lines
can be measured. However, since only a few lines are detected in the UVES
spectra of {\he} available to us in this investigation, equivalent width
measurement errors play a significant role. Furthermore, the lines that
\emph{are} detected are weak, hence errors of $\sim 3$\,m{\AA} have a
comparatively large influence on the abundances. In the last column of Table
\ref{Tab:AbundanceChanges} we summarize the effect that an increase of
$W_{\lambda}$ by $3$\,m{\AA} applied to each individual line in the set of lines
we detected, or to the individual lines we used for deriving upper limits, has
on the derived abundances. With the exception of a very small effect on
\ion{Ca}{2}, due to the large equivalent width of Ca\,{\sc II}~K, the abundance
changes are between 0.06\,dex and 0.13\,dex.

\begin{figure*}[htbp]
  \epsfig{file=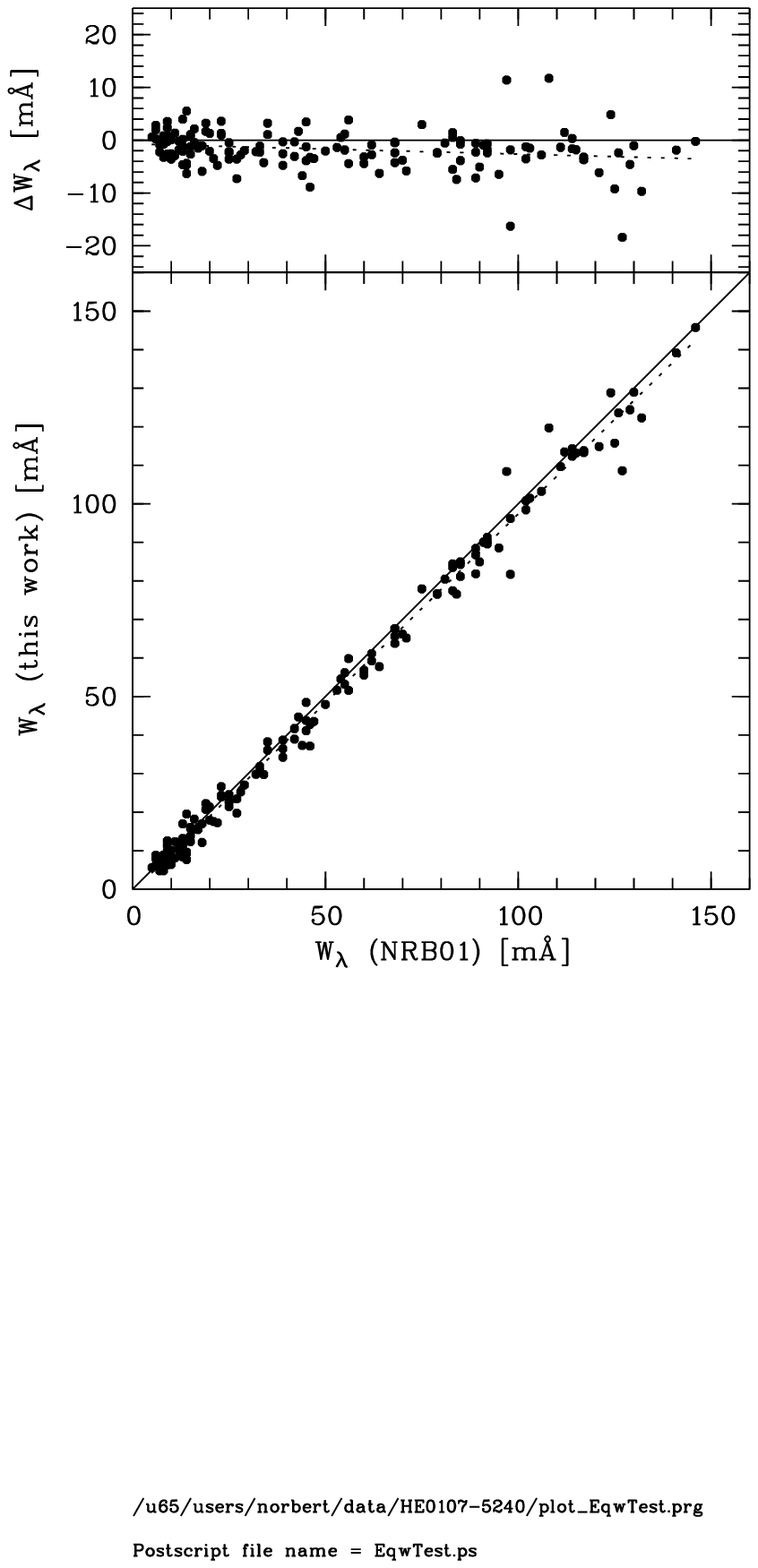, clip=, width=12cm,
    bbllx=73, bblly=283, bburx=314, bbury=599}
  \centering
  \caption{\label{Fig:EqwTest} Equivalent widths of \cd{} measured by
    NRB01, in comparison with our measurements. The solid
    lines indicate a one-to-one correspondence between the measurements; the dotted
    lines are straight line fits to the data. The two sets of measurements
    agree very well; however, our $W_{\lambda}$ measurements possibly yield
    slightly, but systematically, smaller values than those of NRB01. See the text
    for further discussion of this point.  }
\end{figure*}

We applied our equivalent width measurement routine to the UVES spectra of
{\cd}, and compared our results for 145 lines in common with the analysis of
NRB01 with their measurements, kindly made available to us in electronic form
by S.G. Ryan. The agreement between the measurements is very good (see
Figure \ref{Fig:EqwTest}). A straight line fit yields the relation
\begin{equation}
  \label{Eq:EqwNRB01all}
  W_{\lambda}\left(\mbox{us}\right)=-0.71\,\mbox{m{\AA}}+0.98\cdot
  W_{\lambda}\left(\mbox{NRB01}\right).
\end{equation}
The one-sigma scatter is $3.62$\,m{\AA}, in very good agreement with the
expectations from photon statistics, e.g., Equation (\ref{Eq:sigWlambda}).

However, there is an indication that we measure systematically slightly smaller
values than NRB01. This is very likely due to the different method that we
employ for continuum placement. While we perform a $\chi^2$ fit to line-free
wavelength regions, NRB01 determine the continuum level manually.  We note that
the small systematic offset in equivalent widths has only a minor effect on
abundances. For example, in case of \ion{Fe}{1}, converting our measurements to
the scale of NRB01 by means of Equation (\ref{Eq:EqwNRB01all}) leads to a change
of only $+0.03$ dex in [Fe/H], which is negligible compared to other sources of
error (see Table \ref{Tab:AbundanceChanges}, and \S \ref{Sect:AbundanceErrors}
below for a further discussion). A much more accurate continuum placement will
be possible with higher $S/N$ spectra.


\subsubsection{Errors in Physical Data}

Accurately measured $\log gf$ values are available for most species analysed in
{\he}. However, as recently discussed by \citet{KeckpaperII}, comparison of
$\log gf$ values from different sources yields line-to-line scatters as high as
$0.10$--$0.15$\,dex for certain species, e.g., \ion{Fe}{1}, \ion{Fe}{2}, or
\ion{Ti}{2}, while systematic offsets are typically smaller than $0.02$\,dex.

\subsubsection{Errors in Fundamental Stellar Parameters}

The sensitivity of the derived abundances to changes of the stellar parameters
are summarized in Table \ref{Tab:AbundanceChanges}. The dominant error source is
the uncertainty in {\tefft}, resulting in abundance errors of typically
0.1--0.2\,dex. Uncertainties in $\log g$ have the largest effect on abundances
of singly-ionized species, and are negligible for most of the neutral species.
Since the lines we detect in the spectrum of {\he} are all very weak (with the
exception of Ca\,{\sc II}~H and K, and the Balmer lines), the effects of the
microturbulence parameter is negligible. Hence even our large uncertainty in
$v_{\mbox{\scriptsize micr}}$ of 0.5\,km\,$^{-1}$ does not result in significant
abundance errors.

\subsubsection{Errors in Model Atmospheres and in the Spectrum Calculations}\label{ModelatmErrors}

The most important errors in the model atmospheres are likely those due to the
assumption of homogeneous stratification (i.e., the neglect of thermal and
density inhomogeneities), the adoption of mixing-length theory convection, and
those due to the assumption of LTE. The inadequacy of the first-mentioned
assumptions is demonstrated clearly by the 3D hydrodynamical-radiative
simulations of the atmospheres of a Population~II subgiant and dwarf by
\citet{Asplundetal:1999}.  These authors find quite low temperatures in the
upper layers of the models, as compared to 1D MLT models. For spectral lines
partially formed in the outer atmospheric layers (i.e., lines from molecules as
well as low-excitation lines from atoms) one thus would expect weaker lines
predicted by the 1D models than a 3D model would give. This effect is
demonstrated in the work by \citet{Asplundetal:1999}, indicating that 1D models
may yield abundances that are overestimated by typically $0.15$--$0.4$\,dex, and
for molecular lines even by up to 0.6\,dex \citep{Asplund/GarciaPerez:2001}.

The other major uncertainty of the present model, and the subsequent analysis,
is the assumption of LTE. For the model atmosphere, the low metallicity makes
hydrogen, in different forms, such as \ion{H}{1}, H$^-$, H$_2^+$ and
$e^{-}+p^{+}$, the totally dominant species. Of the various processes that are
expected to deviate from LTE, the most important seems to be the ionization of
\ion{H}{1}. The degree of ionization of \ion{H}{1} in the line-forming region is
typically only a factor of 10$^{-5}$, but since this ionization contributes the
electrons for forming H$^-$, the ionization balance is of vital significance for
the atmospheric structure.  Assuming the optically very dense Lyman $\alpha$
transition $n=1\,\rightarrow\, n=2$ to be in detailed balance in most of the
spectrum-forming region, one would believe that photo-ionization from the $n=2$
level is more efficient than in LTE. This is because the radiation from
wavelengths below the Balmer discontinuity comes from deeper atmospheric layers,
unhindered by the heavy metal-line blocking that usually makes this radiation
local in more metal-rich objects. Assessment of the effect that this has on
abundance determinations requires detailed statistical-equilibrium
calculations. However, the tendency will probably be to increase the
photo-ionization and thus the electron pressure, which will increase the H$^-$
absorption and thus the ratios $l_{\nu}/\kappa_{\nu}$ in proportion (here
$l_{\nu}$ denotes the spectral-line absorption coefficient).  This will decrease
calculated equivalent widths correspondingly. If these non-LTE effects are not
considered, the abundances derived will tend to be underestimated.

For the analysis of the spectrum, non-LTE effects for the atoms and molecules
analysed may also be significant. The most important effect will probably be
photoionization, stronger than LTE predicts, again as a result of the hot
radiation fields from deeper photospheric layers.  This is probably only a minor
effect in the determination of abundances from Fe~II, Sr~II or O~I, because
these are ions that are in majority relative to other forms of the element.
These effects may well amount to a few times $0.1$\,dex (see the case of Fe~I
discussed in \S \ref{Sect:FeAbundance}), and again lead to underestimated
abundances.

It is not known to what extent photodissociation of, for instance, CH, will
affect the abundance determinations -- however, underestimates are here also
most probable. It seems that the non-LTE effects for the hydrogen ionization,
affecting H$^-$, and those for metal atoms and other elements, will be in the
same sense and therefore probably cause significant underestimates of the
abundances.

For the determination of many abundances we are in a situation where the effects
of 3D convection and those of departures from LTE may counteract one
another. Both may well be on the order of 0.3\,dex in the derived abundances,
and it is not at all clear to what extent they may cancel, such as they happen
to do for Li in Population~II dwarfs \citep{Asplundetal:2003}. The degree to
which cancellation occurs is also most probably different for different
elements.  It seems that the safest abundance determinations are those based on
ions with abundances that are not very temperature sensitive, such as Ti~II.
Such lines tend to be rather insensitive to both convection uncertainties and
overionization. Although the quality of our UVES spectra are high, we have only
very few such lines detected. Further observations of {\he} with UVES, yielding
spectra of even higher resolution and $S/N$, and covering the whole spectral
range accessible with UVES, from $\sim 3100$\,{\AA} to $\sim 11,000$\,{\AA},
have already been obtained, and will be discussed in future papers (Bessell et
al., in preparation; Christlieb et al., in preparation). Also, a 3D model
atmosphere for {\he} is currently being computed.

\section{DISCUSSION}

\subsection{The Abundance Pattern of \he}\label{Sect:AbundancePattern}

The elemental abundance pattern obtained for {\he} can be summarized 
as follows: 

\begin{itemize}
\item Carbon is very much enhanced, relative to the solar C/Fe, by a factor of
  several thousand to ten thousand.  Nitrogen is also strongly enhanced, by a factor of more
  than one hundred. The abundance ratio C/N is 40--150 . 
\item $^{12}\mbox{C}/^{13}\mbox{C}>50$.
\item There is a significant enhancement of Na, by more than a factor of six,
  while Al does not seem to be enhanced.
\item There is a rather weak enhancement of Mg and possibly of Ca, but not of
  Ti. The upper limit of Si suggests a maximal enhancement of a factor of two
  relative to Fe.
\item The upper limits for Cr and Mn indicate that these elements are
  not enhanced relative to Fe; Ni is slightly depleted, but this is 
  possibly due to non-LTE effects, so that Ni/Fe might actually be close
  to solar.
\item Sr seems to be depleted relative to Fe, while the present upper limits 
  on Ba and Eu are weak, and would allow for enhancements of a factor of 
  $\sim 10$ and $\sim 10^3$, respectively.
\item Upper limits on Sc and Co are about a factor of 10 times solar, while
  that on Zn is as high as $\sim 10^3$ times solar.
\end{itemize}

How do these abundances compare with results obtained for other extremely metal-
poor stars?

A considerable fraction of the giants and subgiants of the extreme Population~II
show strong CH bands and are known to be rich in carbon as well as nitrogen
\citep{Rossietal:1999}. Many of these CH stars also show s-process element
excesses. For example, \citet{Hilletal:2000} have recently analysed two giant
stars, CS~22948-027 and CS~29497-034, with $\mbox{[Fe/H]}=-2.45$ and $-2.90$,
respectively.  These authors found carbon and nitrogen overabundances [C/Fe] and
[N/Fe] of about 2\,dex for the two stars. Sr and Y are enhanced by about 1\,dex,
while Ba, La, Eu and other heavy s- and r-process elements are all enhanced by
about 2\,dex.  While being strongly overabundant in C and N, {\he} does not show
any signs of being rich in s-process elements (as judged from [Sr/Fe] and a poor
upper limit of [Ba/Fe]). However, there are also metal-poor CH stars not
enriched in s-process elements. One well studied giant star of this kind is
CS~22957-027 \citep{Norrisetal:1997b,Bonifacioetal:1998,Preston/Sneden:2001},
which has an effective temperature and a surface gravity close to that of {\he}.
It has [Fe/H] between $-3.0$ and $-3.4$, $\mbox{[C/Fe]}=+2.2$ and also a
considerable N enhancement.  [Na/Fe] is around +0.8\,dex
\citep{Preston/Sneden:2001} and Mg and Ti are moderately enriched. However, Sr
and Ba are depleted relative to Fe and to the Sun by 0.6\,dex or more. Other
similar, but less extreme, giants are CS~30314-067, CS~29502-092, and
CS~22877-001 \citep{Aokietal:2002a}.

It may be tempting to regard CS~22957-027 as a more metal-rich version of
{\he}. However, Preston \& Sneden found that CS~22957-027 is a spectroscopic
binary with a period of about 3125 days. The binary character of the star was
not expected, since its low values of the s-process elements does not suggest
its atmospheric composition to be the result of mass transfer from a companion
in the Asymptotic Giant Branch (AGB) stage. Preston \& Sneden speculated that
this star has developed into a carbon star as a result of internal mixing after
the first He shell flashes \citep[case II of][]{Fujimotoetal:2000}, and that it
just happens to have a companion.

Another star, which may have a chemical composition of the surface layers
similar to that of {\he}, is the dwarf carbon star G77-61. It was found by
\citet{Gassetal:1988} to have a chemical composition with $\log (\epsilon)$ of
7.3, 5.2, and 2.0 for C, N, and Fe, respectively. The corresponding values for
{\he} are 7.0, 4.8 and 2.2. However, Gass et al. found abundances of Na and Mg
that are typically a factor of 10 higher than our values for {\he}. G77-61 also
shows velocity shifts of its spectral lines with a period of $P=245$\,days,
which are ascribed to a companion \citep{Dearbornetal:1986}. Concerning the
spectral analysis, we note that the complex spectrum of G77-61 is difficult to
analyse, and that the quality of the spectrum used by \citet{Gassetal:1988} was
limited (i.e., $R\sim 20,000$; $S/N=30$ per resolution element). The possible
similarities between the abundance pattern of G77-61 and {\he}, as well as the
low iron abundance claimed for the former star by Gass~et~al., require
confirmation by an analysis based on higher quality data before any conclusions
can be drawn. Such analyses, based on Subaru/HDS and Keck/HIRES spectra are in
progress (Aoki, private communication; Plez \& Cohen, private communication).

No significant radial velocity shifts have been found for {\he} so far.  In this
respect it seems to depart from both CS~22957-027 and G77-61, but both binaries
are long-period, low-amplitude systems.  If {\he} is a binary with a period as
long as either of these two stars, it probably would not have been detected
based on the radial velocity measurements made to date.

The comparison with other extremely metal-poor stars suggests that the abundance
pattern of {\he} is not totally unique, although the star is certainly much more
iron-poor than any other giant discovered until now. With this background, we
discuss the astrophysical origin of the abundance pattern of this star below.

\subsection{Nucleosynthesis and {\he}}

The observed abundance pattern raises the following main questions:
\begin{enumerate}
\item[(a)] What were the processes, and what were the sites, where the heavy
nuclei of this star were formed?
\item[(b)] How were these nuclei acquired by this star? Did they arise due to
enrichment from earlier supernovae (SNe), or other objects, or from the previously
polluted insterstellar cloud in which the star was formed? Did they arise via
accretion from a neighboring star (presumably a binary companion)? Did they
arise via accretion from the general interstellar medium (ISM) after multiple
passages of the star through the Galactic plane? Did they arise as a result of
dredging up processed material from the deeper layers of the star itself?
\item[(c)] Could the surface composition of the star have been affected by
  other processes, such as diffusion or selective radiative pressure
  expulsion?
\end{enumerate}

Below we address these issues, guided by contemporary calculations of the
evolution of very metal-poor stars and by calculations of yields from
SNe and from other contributors of heavy elements in the early Galaxy.

\subsubsection{Synthesis in Supernovae}

\begin{figure*}[p]
  \begin{center}
  \epsfig{file=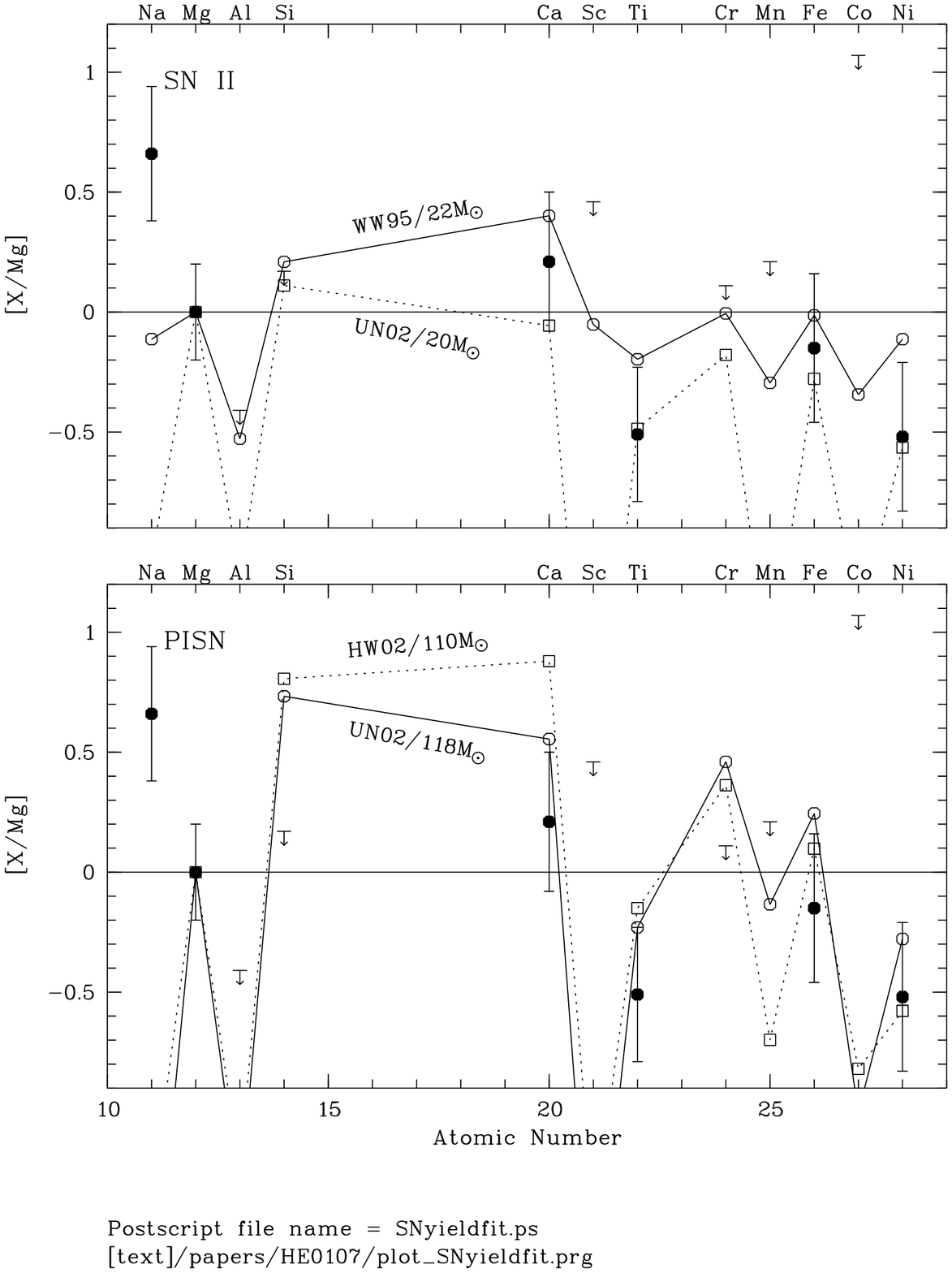, clip=, width=14cm,
    bbllx=44, bblly=139, bburx=526, bbury=722}
  \end{center}
  \centering
  \caption{\label{Fig:SNyieldfit} Comparison of the observed abundance pattern 
  of {\he} with the yields of SN~II (upper panel) and pair-instability supernovae
  (lower panel). Filled circles and arrows denote abundances measured for
  {\he}, and upper limits, respectively. Note that the latter are plotted
  without error margin. The abundances were normalized relative to the Mg
  abundance. In the lower panel, masses refer to He core masses.
  HW02 = \citet{Heger/Woosley:2002}, UN02 = \citet{Umeda/Nomoto:2002},
  WW95 = \citet{Woosley/Weaver:1995}.}
\end{figure*}

The relative abundances of {\he} might have been created by yields from a SN
with $Z_{\mbox{\scriptsize init}}=0.0$ and a mass of about 15--25 solar masses,
as modeled by \citet{Woosley/Weaver:1995} and \citet{Umeda/Nomoto:2002}. This is
illustrated in Figure \ref{Fig:SNyieldfit}. The yields of massive Pop. III
pair-instability supernovae (i.e., hypernovae) computed by
\citet{Heger/Woosley:2002} or \citet{Umeda/Nomoto:2002} also offer rather good
fits to our observed abundances. However, all of the above mentioned supernova
models do not at all explain the high carbon, nitrogen, and sodium
abundances. The predicted amounts of C and N depart from those observed in {\he}, by
typically 4--5\,dex.

\citet{Norrisetal:2002}, in an attempt to explain the chemical composition of
 CS~22949-037 ($\mbox{[Fe/H]}=-3.8$, $\mbox{[C/Fe]}=+1.05$, $\mbox{[N/Fe]}=
 +2.3$), refer to the study of massive non-rotating and rotating
 pair-instability SN models by \citet{Fryeretal:2001}. The latter authors
 suggested that in the case of pair-instability SN the shear between the
 convective hydrogen shell and the core can become large enough to lead to a
 significant dredge-up of the helium core, i.e. helium and large amounts of
 helium-burning products (carbon, oxygen, neon) are mixed into the envelope.
 Some of the carbon and oxygen are then burned by the CNO process to
 nitrogen. Fryer et al. found total envelope masses of carbon and nitrogen of
 $0.65$ and $9.48$ solar masses, respectively, for a 250\,M$_{\odot}$ model,
 while the corresponding envelope masses for a $300$\,M$_{\odot}$ model are
 $0.09$ solar masses of carbon, $1.56$ of nitrogen, $17$ of oxygen and $2.9$ of
 neon and $1.10$ of magnesium.  The latter model does not explode but turns into
 a black hole.  However, as discussed by Norris et al., it may be possible that
 at least some of the envelope could be expelled due to a stellar wind,
 pulsations, etc. If the nitrogen and a significant fraction of the carbon were
 produced by such hypernovae, an important problem will be to limit the yield of
 heavier elements like Mg.

\citet{Umeda/Nomoto:2003} proposed that the abundance pattern of {\he} arises
from material that has been enriched by a $25\,M_{\odot}$ Population~III star
exploding as a supernova of low explosion energy ($E_{\mbox{\scriptsize
exp}}=3\cdot 10^{50}\,\mbox{erg}$). By assuming that the material produced
during the SN event is homogeneously mixed over a wide range of the mass
coordinate, and a large fraction of the material falls back onto the compact
remnant (the ``mixing and fallback'' mechanism), Umeda \& Nomoto are able to
reproduce the abundance pattern of {\he} quite well (see their Figure
1). In particular, very high C/Fe and N/Fe ratios can be produced in this
scenario, with the CNO elements produced in late stages of the evolution of the
SN progenitor \citep[for a similar work on the abundance patterns of
CS~29498-043 and CS~22949-037 see][]{Tsujimoto/Shigeyama:2003}.

\subsubsection{Synthesis in Red Giants}

The strong overabundance of carbon may suggest {\he} to be a carbon star,
i.e. a star that has produced its own $^{12}$C by He burning and subsequent
mixing out of the processed material, presumably in a He shell flash during the
AGB stage of evolution.  However, a strong argument against this possibility is
provided by the derived surface gravity, which is higher than those
characteristic of AGB stars (which typically have $\log g < 1.0$ in cgs units).
An alternative to this explanation would then be that the star was polluted by a
more massive companion that had previously evolved into a carbon star.

These alternatives obviously assume two different nucleosynthesis sites -- a red
giant origin of the CN(O) elements and of some of the Na and Mg, and an earlier
supernovae origin for the rest of the elements.

\citet{Fujimotoetal:2000} discuss the origin of extremely metal-poor carbon
stars. In their evolutionary low-mass ($\mbox{M} < 1.0\,\mbox{M}_{\odot}$)
models for extremely metal-poor stars ($\mbox{[Fe/H]}<-4$), the helium
convection during the first off-centre helium core flash extends into the
hydrogen-containing layers. This brings hydrogen down to hotter regions, leading
to intensive hydrogen burning (a ``H flash''), and matter that has experienced
helium-burning reactions and then is further processed by hydrogen burning
reactions, is subsequently brought to the surface. As a result, for a
0.8\,M$_{\odot}$ model, a surface abundance close to solar (relative to
hydrogen) would result for carbon. Roughly equal amounts of nitrogen would also
be produced by subsequent CNO burning. For a somewhat more massive model,
$\mbox{M}=0.9$\,M$_{\odot}$, a similar process occurs. On the other hand, for
models with $\mbox{M} \geq 1.0\,\mbox{M}_{\odot}$, helium ignites in the core
before the electrons have become degenerate and before a temperature inversion
in the core has been established due to neutrino losses.  For somewhat more
metal-rich models ($\mbox{[Fe/H]}>-4$), Fujimoto et al. find that the outer edge
of the core-flash driven convection shell barely touches the hydrogen-rich
matter, and further mixing does not occur during the core flash.  Later,
however, after the first shell flashes at the base of the AGB, the surface
convection zone extends down into layers formerly occupied by the
helium-convection shell, and for a $\mbox{M}=0.8\,\mbox{M}_{\odot}$ model again
a roughly solar carbon surface abundance is produced, while the nitrogen
abundance is lower by one order of magnitude.  This higher C/N ratio (about 6
instead of 1) is due to the larger carbon abundance in the helium convection
zone for this case.

\citet{Fujimotoetal:2000} argue that no third dredge-up with heavy elements
produced by the s-process should occur for stars with $\mbox{[Fe/H]}<-2$ and
$\mbox{M}<1.0\,\mbox{M}_{\odot}$. Thus, both for the most metal-poor case, when
the He core flash produces the carbon enrichment, and for the more metal-rich
case when the first shell flashes on the AGB are responsible, the
s-process-element enrichment can be avoided.

\citet{Schlattletal:2002} carried out similar model calculations and verified
several of the findings by Fujimoto et al. However, they find the occurance of
the H-flash to be quite sensitive to details, such as the exact He abundance and
the assumptions concerning diffusion, in particular for higher masses (around
1\,M$_{\odot}$). They also find resulting carbon and nitrogen abundances
relative to iron of about 4 orders of magnitude greater than those of the Sun,
and $^{12}$C/$^{13}$C ratios of about 5. They discuss the fact that their models
produce far too much carbon and nitrogen, as compared with carbon-rich
Population~II giants such as CS~22892-052, and presume that this failure might
be due to the crude one-dimensional description used to calculate the violent H
burning in the helium-flash driven convective zone.

\citet{Siessetal:2002} followed the evolution of models of metal-free stars up
the AGB with masses from 0.8--20 solar masses, and found in the 1--5 solar mass
models that a secondary convective zone developed at the He-H discontinuity in
the beginning of the AGB phase.  This region expands and engulfs gas that just
has been carbon-enriched in the He-burning shell, as well as hydrogen from upper
layers. H-burning flashes occur with CNO burning, and subsequently the
convective envelope dredges the products to the surface.  The envelopes thus
become considerably enriched by $^{12}$C and $^{14}$N, as well as $^{23}$Na and
$^{25,26}$Mg. For instance, their 1.0\,M$_{\odot}$ model ends with a surface
abundance of carbon of 3 times solar, while the C/N ratio is about 10. This
seems to suggest that the C and N enrichment of {\he} could well be the result
of mass transfer from such a companion. The models of Siess et al. also suggest
an enrichment of Na and Mg, with two orders of magnitude greater Na/C and Mg/C
ratios than those observed {\he}. Also, some s-process elements may possibly be
produced in these models, despite the lack of iron seeds.  However, the
predicted amounts of Na, Mg and the s-process elements are relatively uncertain.
The hypothesis of the origin of carbon and nitrogen as a result of
accretion from an AGB companion following this path still seems possible.


From the work by Fujimoto et al., Schlattl et al., and Siess et al. we conclude
that {\he} may have been polluted by a low-mass companion that evolved up to the
tip of the red-giant branch, and produced considerable amounts of C and N as a
result of the He core flash. This material was later deposited onto our star.
The arguments against this hypothesis come from the predicted low C/N ratio, as
compared with the ratio observed, and the high $^{12}$C/$^{13}$C ratio of the
star.  Another hypothesis from this point of view might be an evolution of the
star itself, similar to that found by Fujimoto et al. for models with
$-4<\mbox{[Fe/H]}<-2$, where the surface carbon is produced by the first He
shell flashes, at the base of the AGB, leading to higher C/N ratios.  In this
case, the star could show the CNO abundance peculiarities already at the base of
the AGB. However, the radius of the star at this stage should already be so
large that it is barely compatible with our lower limit of the surface gravity.

\subsection{Other Hypotheses: Selective Dust Depletion, Accretion from the ISM}

\begin{figure*}[htbp]
  \begin{center}
  \epsfig{file=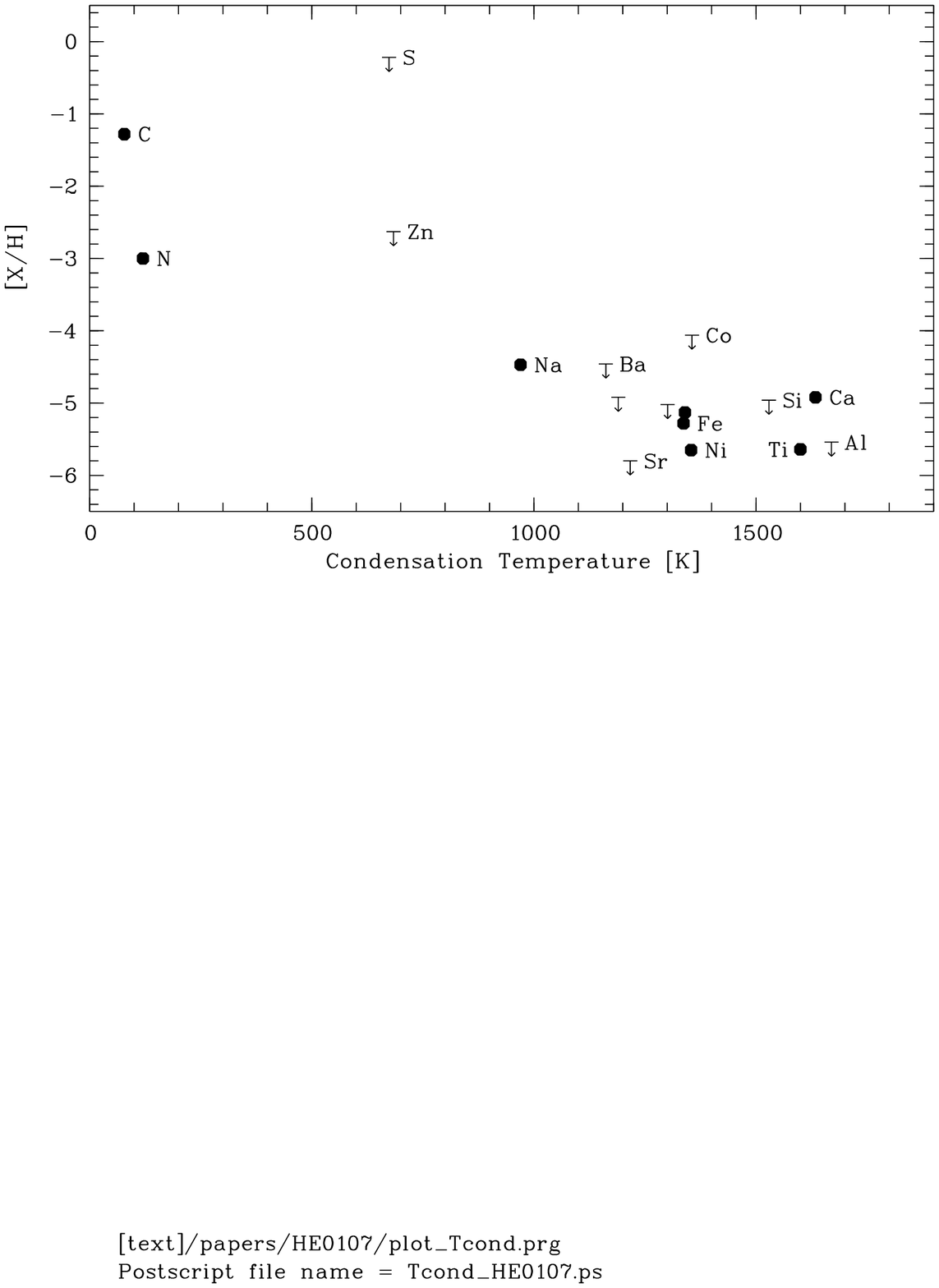, clip=, width=14cm,
    bbllx=52, bblly=419, bburx=526, bbury=712}
  \end{center}
  \centering
  \caption{\label{Fig:Tcond_HE0107} Elemental abundances (filled circles) and
     upper limits (arrows) measured in {\he} versus dust condensation
     temperatures, $T_c$. The values for $T_c$ are taken from
     \citet{Lodders/Fegley:1998}.}
\end{figure*}

In \citet{HE0107_Nature}, we considered the possibility that {\he} is a
Population~II post-AGB star, similar, e.g., to RV~Tauri stars of Preston type B
or C \citep[][ and references therein]{Giridhar:2000}. These pulsating
metal-poor stars are generally believed to be affected by dust-gas separation,
which may systematically deplete the atmospheres of elements that may condense
at high temperature. In fact, plotting the abundances of the different elements
versus dust condensation temperatures, we find a possible strong correlation in
this direction (see Figure \ref{Fig:Tcond_HE0107}), as is also seen in
metal-poor post-AGB stars \citep[see, e.g.,][]{Tramsetal:1993}. However, the
existence of a correlation is only supported by the high values of the CN
abundances, with this interpretation then suggesting original abundances of the
star corresponding to $\mbox{[Fe/H]}\sim -1.4$. The correlation is also
consistent with our upper limit on the Zn abundance. Another abundance that
could support the hypothesis would be that of sulphur, with a relatively low
condensation temperature. Unfortunately, it is not possible to derive a
meaningful upper limit for S from the spectra used in this study . On the other
hand, the high C/N ratio observed in {\he} is not observed in other metal-poor
AGB stars \citep{Tramsetal:1993}, and cannot be explained by selective
depletion.

The similarity of {\he} with the RV~Tauri stars is mainly restricted to its
low metal abundances as compared with its CN abundances. Its effective
temperature is at the lower end of the temperature interval of observed stars
of type B and C \citep[see][]{Giridharetal:2000}. The steep increase of the
mass of the convective zone when the temperature decreases as a star approaches
the giant branch suggests that any chemical inhomogeneities at the surface of
{\he} should be diminished by mixing. Also, the star is neither known to be
pulsating, nor a binary, which the typical, strongly affected RV~Tauri stars
tend to be \citep{VanWinckeletal:1995,VanWinckeletal:1999}. We therefore
conclude again that the chemical peculiarities of {\he} are probably not the
result of selective dust condensation of the type that has been active in
RV~Tauri stars, although further efforts to measure lines of Zn and S should
be made in order to exclude, or potentially open up, this possibility.

The abundances of most heavy elements observed in {\he} are so low, and the age
of the star presumably so large (on the order of the age of the Universe), that
one should consider the possibility that its abundances have been severely
changed by accretion from the interstellar medium. It may even have been formed
out of totally metal-free gas and accreted its heavy nuclei during repeated
passages through the Galactic disk.  The possibility that halo stars accrete
metal-rich material through encounters with interstellar clouds in the Galactic
plane was suggested and discussed by \citet{Talbot/Newman:1977},
\citet{Yoshii:1981}, and \citet{Iben:1983}. Yoshii made calculations of the
accretion rates onto halo stars, and found typical amounts of accreted material
onto a 0.8\,M$_{\odot}$ halo star, from the formation of the Galactic disk to
present times, of $10^{-5}$ to $10^{-2}$ solar masses, depending on the adopted
orbital parameters. Following \citet{Yoshii:1981} and adopting a mass of the
convective envelope of {\he} of 0.2--0.3\,M$_{\odot}$ and a solar composition of
the Galactic gas disk, we find that accretion should lead to heavy-element
abundances in the range $-4.5<\mbox{[X/H]}<-1.3$. As judged from this estimate,
all heavy nuclei of our star could be accreted.  However, Yoshii's estimate is
still uncertain. For example, the effects of a stellar wind ``protecting'' the
star from accretion are not considered.

A variation of the accretion scenario is discussed by
\citet{Shigeyamaetal:2003}. They estimate that considerable accretion rates (of
the order of $10^{-13}\mbox{M}_{\odot}\mbox{yr}^{-1}$) can be reached if the
velocity of the star relative to the gas cloud from which it accretes is low
(i.e., $\sim 10$\,km/s). The solar mass-loss rate is $\sim
10^{-14}\mbox{M}_{\odot}\mbox{yr}^{-1}$, and the authors also argue that the
wind of an extremely metal-deficient dwarf is expected to be much weaker than
that of the Sun.

We conclude that most heavy elements of {\he} {\it might} be accreted from the
interstellar medium, although this is most probably not the case for C, N, and
Na.


\section{DISCUSSION AND CONCLUSIONS}\label{Sect:Conclusions}

We have found an overall metal abundance of {\he}, disregarding the CNO
elements, of less than $10^{-5}$ solar, or more than one order of magnitude less
than any other known extremely metal-poor giant star. It is tempting to
speculate that the chemical composition of this star might reflect
nucleosynthesis processes that occured during the very first star formation
period of the universe, only about 200\,Myr from the Big Bang, according to
recent WMAP data \citep{Bennettetal:2003b, Kogutetal:2003}. To what extent this
is the case, and for which elements, remains to be explored.  Obviously, an
important issue to be investigated in the future is the similarity of the
abundance pattern of {\he} with other stars with $\mbox{[Fe/H]}<-4.0$, if such
stars exist.

From the discussion above we conclude that most elements heavier than Na in
{\he} may be the result of one or a few supernova explosions. We cannot disprove
the alternative hypothesis that these elements were gradually accreted e.g. from
the interstellar medium in the Galactic disc during passages of the star through
it.

Concerning the origin of carbon and nitrogen, one hypothesis is that the star
has made them through helium and hydrogen burning and mixed them to the surface
as the result of a He and H flash. Also, Na may have been formed this way.  If
this self-enrichment scenario is true, a problem remains -- why are the observed
amounts of C and N significantly smaller than those predicted by the models?
One way around this would be to assume that {\he} was instead polluted by a
binary companion that went through this development.  In this case there would
still be a problem with the C/N ratio, which is predicted by current models to
be one order of magnitude greater than the observed one, as well as the
$^{12}$C/$^{13}$C ratio, which is predicted to be much smaller. A hypothetical
companion probably also went through a number of He flashes which might have
produced some s-process elements. More precise measurements of the abundance of
the s-process elements, as well as oxygen, and radial velocity measurements
spanning a longer period of time, will put stronger constraints on possible
scenarios for the origin of C and N in the atmosphere of {\he}.

As already discussed in \citet{HE0107_Nature}, it has been thought, until very
recently, that with the exception of the metal-depleted post-AGB stars (and
G77--61, for which no accurate Fe abundance measurement is available), stars of
$\mbox{[Fe/H]}<-4.0$ did not exist. This observational limit is based on the
failure of even very extensive wide-angle spectroscopic surveys, like the HK
survey of Beers, Preston and Shectman \citep{BPSII,TimTSS}, to reveal a star
more metal-poor than {\cd}. This failure is easily explainable theoretically,
e.g., by the fact that a massive star of, say, $15\,\mbox{M}_{\odot}$ has an
evolutionary timescale that is about one order of magnitude shorter than the
contraction timescale of a $0.8\,\mbox{M}_{\odot}$ star. It was therefore
thought that the massive members of the first generation of stars in our Galaxy,
ending up as supernovae of type II (SN~II), might have enriched the ISM to
$\mbox{[Fe/H]}>-4.0$ {\it before} low-mass stars formed. Alternatively, low-mass
star formation might have been suppressed by less efficient cooling due to the
absence of metals \citep[e.g.,][]{Bond:1981}. The discovery of {\he} can be a
challenge to these scenarios.

However, if the first generation of stars was predominantly located in the mass
range $20$--$130$\,M$_{\odot}$, and exploded as low-explosion-energy supernovae
as suggested by \citet{Umeda/Nomoto:2003}, the influence of the presence of C
and O in nearly primordial gas clouds on cooling needs to be taken into account.
Umeda \& Nomoto report that ``the cooling efficiency is large enough to form
small mass stars'' in this case.


One prediction of the model of Umeda \& Nomoto is that {\he} should have an
oxygen abundance of $\mbox{[O/Fe]}\sim 2.8$\,dex. We shall confront this
prediction with more recent observations in a forthcoming paper.

A consequence of the scenario of Umeda \& Nomoto would be that, if there is a
critical overall metallicity, $Z_{\mbox{\scriptsize crit}}$, which needs to be
reached in order to make the formation of low-mass stars possible, \emph{all}
low-mass stars below $\mbox{[Fe/H]} < \log\left(Z_{\mbox{\scriptsize
crit}}/Z_{\odot}\right)$ should be enhanced in CNO relative to Fe and other elements,
giving rise to more efficient cooling of the gas, and the subsequent enhanced
formation of such stars.  This scenario could be tested by the assemblage of a
significant sample of metal-poor stars, extending to $\mbox{[Fe/H]}<-4.0$, and
determination of the fraction of carbon-enhanced stars in that sample.

We note that the Umeda \& Nomoto scenario is also attractive in that it
does not require the existence of a binary companion for the carbon-enhanced
stars. In fact, there are at least three unevolved carbon-enhanced metal-poor
stars known that do not show any radial velocity variations larger than
0.4\,km\,s$^{-1}$ over a period of 8 years \citep{Preston/Sneden:2001}, ruling
out, at least for these three stars, the scenario in which a formerly more
massive star went through its AGB phase and transfered dredged-up material onto
the surface of the less-massive companion.

It should be realized, when considering the attractive features of the Umeda \&
Nomoto model, that it, unlike the other models discussed, was proposed \emph{a
posteriori}, with knowledge about the peculiarities of {\he}. At present, we
consider it premature to judge whether this model is to be preferred to the SN
plus red giant alternative.

{\he} was found in a sample of roughly $300$ stars with $\mbox{[Fe/H]}<-3.0$
identified in the Hamburg/ESO and HK surveys. Assuming that {\he} is not unique, and
considering that up to now only roughly $1/3$ of the HES metal-poor candidates
have been followed-up on, it is not excluded that one or more additional stars
with $\mbox{[Fe/H]}<-4.0$ will be found in the HES, or other proposed surveys in
the near future.

\acknowledgments

We would like to express our gratitude to the ESO staff on Paranal and Garching
for carrying out the observations with VLT-UT2, and reducing the data,
respectively. Allocation of observing time at the VLT by the Director's
Discretionary Time committee is acknowledged. Exploitation of the stellar
content of the HES would have been impossible without the contributions of
L. Wisotzki and D. Reimers. We thank M. Asplund, B.  Edvardsson, J. Lattanzio,
J. Norris, N. Piskunov, B. Plez, D. Reimers, S.G. Ryan and L. Siess for
discussions. N. Piskunov and E. Stempels patiently taught N.C. how to use
\texttt{REDUCE} and related tools. We are grateful to M. Skrutskie for
supplying 2MASS results in advance of release. We thank P. Wood and M. Dehn for
obtaining $JHK$ and Str\"omgren $uvby$ photometry, respectively. CH and CN line
lists were kindly provided by B. Plez. We appreciate valuable comments of
D. Lambert, and thank him for forwarding a table with dust condensation
temperatures to us. Careful proof-reading by A. Frebel is gratefully
acknowledged. This research made extensive use of the Vienna Atomic Line
Database (VALD), and the Abstract Service of NASA's Astrophysics Data
System. N.C. acknowledges financial support through a Marie Curie Fellowship of
the European Community program \emph{Improving Human Research Potential and the
Socio-Economic Knowledge} under contract number HPMF-CT-2002-01437, and from
Deutsche Forschungsgemeinschaft under grant Re~353/44-1. T.C.B. acknowledges
financial support from the U.S. National Science Foundation under grants
AST~00-98508 and AST~00-98549, and he thanks the Hamburger Sternwarte for
hospitality shown during his visit when selection and visual inspection of the
HES EMP candidates took place, among which {\he} was found. We also thank Silvia
Rossi for her help in this process. The Uppsala group acknowledges support from
the Swedish Research Council (VR).


\bibliography{HES,add,atomdata,carbonstars,datared,instr,ncastro,ncpublications,mphs,modelatm,photometry}
\bibliographystyle{apj}

\clearpage

\begin{deluxetable}{lllr} 
\tablecolumns{4} 
\tablewidth{0pt} 
\tablecaption{\label{Tab:UVESobs} UVES observations of HE~0107$-$5240 and
  CD~$-38^{\circ}\,245$} 
\tablehead{
  \colhead{} & \colhead{} & \colhead{} & \colhead{$t$}\\
  \colhead{Target} & \colhead{UT date\tablenotemark{a}} &
  \colhead{UT\tablenotemark{a}} & \colhead{(sec)}
  }
\startdata 
 HE~0107$-$5240        & 20-Dec-2001 & 00:55:45 &  3600\\
 HE~0107$-$5240        & 20-Dec-2001 & 02:03:31 &  3600\\
 HE~0107$-$5240        & 20-Dec-2001 & 03:04:58 &  3600\\
 HE~0107$-$5240        & 21-Dec-2001 & 02:22:20 &  3600\\
 HE~0107$-$5240        & 21-Dec-2001 & 01:24:52 &  3000\\
 CD~$-38^{\circ}\,245$ & 21-Dec-2001 & 01:01:26 &   900\\
\enddata
\tablenotetext{a}{At beginning of observation}
\end{deluxetable} 

\begin{deluxetable}{lrrrrrrrrr} 
\tablecolumns{10} 
\tablewidth{0pt}
\tablecaption{\label{Tab:Photometry} Broad- and Intermediate-Band Photometry of
  HE~0107$-$5240.} 
\tablehead{
  \colhead{} & \colhead{$c_1$} & \colhead{$b-y$} &  \colhead{$B$} & \colhead{$V$} & 
  \colhead{$R$} & \colhead{$I$} & \colhead{$J$} & \colhead{$H$} & \colhead{$K$} \\
  \colhead{Run} & \colhead{(mag)} & \colhead{(mag)} & \colhead{(mag)} & \colhead{(mag)} &
  \colhead{(mag)} & \colhead{(mag)} & \colhead{(mag)} & \colhead{(mag)} & \colhead{(mag)}
  }
\startdata 
ESO-DK 1.54\,m\tablenotemark{a} & \nodata & \nodata &   15.86 &   15.18 &   14.73 &   14.29 & \nodata & \nodata & \nodata \\
LCO 1\,m\tablenotemark{b}       &    0.09 &    0.50 & \nodata &   15.16 & \nodata & \nodata & \nodata & \nodata & \nodata \\
2MASS/JG\tablenotemark{c}       & \nodata & \nodata & \nodata & \nodata & \nodata & \nodata &   13.70 &   13.20 &   13.24 \\
SSO 2.3\,m\tablenotemark{d}     & \nodata & \nodata & \nodata & \nodata & \nodata & \nodata &   13.68 &   13.23 &   13.22 \\
Adopted                         &    0.09 &    0.50 &   15.86 &   15.17 &   14.73 &   14.29 &   13.69 &   13.22 &   13.23 \\
\enddata
\tablenotetext{a}{ESO-Danish 1.54\,m/DFOSC, December 2001, observer: T.C. Beers}
\tablenotetext{b}{Las Campanas 1\,m/Direct CCD Camera, 30 December 2001, observer: M. Dehn}
\tablenotetext{c}{2MASS values transformed to the Johnson-Glass system of
   \citet{Bessell/Brett:1988}, using the transformations listed in \citet{Carpenter:2001}.}
\tablenotetext{d}{SSO 2.3\,m/CASPIR, 22 December 2001, observer: P. Wood} 
\end{deluxetable} 

\begin{deluxetable}{lrlcrlrl} 
\tablecolumns{8} 
\tablewidth{0pt} 
\tablecaption{\label{Tab:TeffDerivation} Derivation of {\tefft} for
  HE~0107$-$5240 and CD~$-38^{\circ}\,245$.} 
\tablehead{
  \colhead{} & \multicolumn{2}{c}{{\he}} & \colhead{} & \multicolumn{2}{c}{{\cd}} & \colhead{}\\
  \cline{2-3} \cline{5-6} \\ 
  \colhead{Measured} & \colhead{Value} & \colhead{Derived {\tefft}} & \colhead{}
  & \colhead{Value} & \colhead{Derived {\tefft}} & \colhead{$\Delta\teffm$} & \colhead{Notes}\\
  \colhead{quantity} & \colhead{}      & \colhead{(K)}              & \colhead{}
  & \colhead{} & \colhead{(K)} & \colhead{(K)} & \colhead{}
  }
\startdata
 H$\alpha$ profile                    & \nodata   & $5140\pm 200$ & & \nodata       & $4710\pm 150$    & $430$ & STEHLE$+$BPO\\ 
 H$\alpha$ profile                    & \nodata   & $5180\pm 150$ & & \nodata       & $4800\pm 200$          & $380$ & VCS$+$AG\\ 
 $(B-V)_{0}$/J    & 0.68\,mag & $5150\pm\phn 90$ & & 0.805\,mag\tablenotemark{a}    & $4880\pm\phn 50$ & $270$ & HBS00\\
 $(V-R)_{0}$/C    & 0.43\,mag & $5190\pm 170$    & & 0.497\,mag\tablenotemark{a}    & $4860\pm\phn 70$ & $330$ & HBS00\\
 $(V-I)_{0}$/C    & 0.86\,mag & $5210\pm 100$    & & 1.010\,mag\tablenotemark{b}    & $4860\pm\phn 50$ & $350$ & HBS00\\
 $(V-K)_{0}$/JG   & 1.90\,mag & $5320\pm\phn 60$ & & 2.33\phn\,mag\tablenotemark{c} & $4840\pm\phn 50$ & $480$ & HBS00\\
 $(B-V)_{0}$/J    & 0.68\,mag & $5080\pm 190$    & & 0.805\,mag\tablenotemark{a}    & $4780\pm 100$    & $300$ & AAM99b \\
 $(V-K)_{0}$/JTCS & 1.95\,mag & $5210\pm\phn 60$ & & 2.38\phn\,mag\tablenotemark{d} & $4670\pm\phn 40$ & $540$ & AAM99b\\
 $(b-y)_0$        & 0.49\,mag & $5100\pm 150$    & & 0.594\,mag\tablenotemark{b}    & $4800\pm\phn 70$ & $300$ & AAM99b\\
 $(b-y)_0$        & 0.49\,mag & $5080\pm 250$    & & 0.594\,mag\tablenotemark{b}    & $4730\pm 130$    & $350$ & Clem98\\
\enddata
\tablecomments{Colors with qualifier `J', `C', `JG' refer to the
  Johnson, Cousins, Johnson-Glass system, respectively. `JTCS' refers to
  $V-K$ colors where $K$ has been converted to the TCS system, using
  the transformation listed in \citet{Alonsoetal:1998}.
  }
\tablenotetext{a}{\citet{Bessell/Norris:1984}}
\tablenotetext{b}{\citet{Mermilliodetal:1997}}
\tablenotetext{c}{Peterson et al. (1990)}
\tablenotetext{d}{Assuming that the transformation of the $K$ magnitude to the
   TCS system leads to the same correction as for {\he}, i.e., assuming that
   {\he} and {\cd} have similar $J-K$ colors. }
\tablerefs{HBS00: \citet{Houdasheltetal:2000}; AAM99b: \citet{Alonsoetal:1999b,Alonsoetal:2001};
   Clem98: \cite{Clem:1998}.
}
\end{deluxetable}

\begin{deluxetable}{lll} 
\tablecolumns{2} 
\tablewidth{0pt} 
\tablecaption{\label{Tab:StellarParameters} Stellar parameters adopted
  for HE~0107$-$5240.} 
\tablehead{
  \colhead{Parameter} & \colhead{Value} & \colhead{$\sigma$}
  }
\startdata 
 $T_{\mbox{\scriptsize eff}}$  & $5100$\,K & $150$\,K\\
 $\log g$ (cgs)                & $2.2$\,dex  & $0.3$\,dex\\
 $\mbox{[Fe/H]}$               & $-5.3$\,dex & $0.2$\,dex \\
 $v_{\mbox{\scriptsize micr}}$ & $2.2$\,km\,s$^{-1}$ & $0.5$\,km\,s$^{-1}$
\enddata
\end{deluxetable} 

\begin{deluxetable}{lllrrr} 
\tablecolumns{6} 
\tablewidth{0pt} 
\tablecaption{\label{Tab:Linelist} Equivalent widths, upper limits,
  and line-by-line abundances for HE~0107$-$5240}
\tablehead{
  \colhead{}    & \colhead{$\lambda$} & \colhead{$\chi$} & \colhead{$\log gf$} &
  \colhead{$W_{\lambda}$} & \colhead{$\log\epsilon$} \\
  \colhead{Ion} & \colhead{({\AA})}    & \colhead{(eV)}  & \colhead{(dex)}    &
  \colhead{(m{\AA})} & \colhead{(dex)}
  }
\startdata
\ion{Li}{1} & $6707.761$ & $0.00$ & $-0.01$ & $<11$ & $<1.12$ \\
\ion{Li}{1} & $6707.912$ & $0.00$ & $-0.31$ & $<11$ & $<1.42$ \\
\ion{Na}{1} & $5889.951$ & $0.00$ & $ 0.12$ & $ 30$ & $ 1.83$ \\
\ion{Na}{1} & $5895.924$ & $0.00$ & $-0.18$ & $ 19$ & $ 1.90$ \\
\ion{Mg}{1} & $5172.684$ & $2.71$ & $-0.40$ & $ 10$ & $ 2.38$ \\
\ion{Mg}{1} & $5183.604$ & $2.72$ & $-0.18$ & $ 18$ & $ 2.45$ \\
\ion{Al}{1} & $3961.520$ & $0.01$ & $-0.32$ & $<10$ & $<0.93$ \\
\ion{Si}{1} & $3905.523$ & $1.91$ & $-1.09$ & $<20$ & $<2.55$ \\
\ion{S}{1}  & $4034.028$ & $6.86$ & $-1.69$ & $<13$ & $<7.11$ \\
\ion{Ca}{1} & $4226.728$ & $0.00$ & $ 0.24$ & $ 25$ & $ 0.99$ \\
\ion{Ca}{2} & $3933.663$ & $0.00$ & $ 0.14$ & \nodata & $ 1.44$ \\
\ion{Sc}{2} & $3572.526$ & $0.02$ & $ 0.27$ & $<30$ & $<-1.22$ \\
\ion{Sc}{2} & $3613.829$ & $0.02$ & $ 0.42$ & $<25$ & $<-1.50$ \\
\ion{Sc}{2} & $4314.083$ & $0.62$ & $-0.10$ & $<10$ & $<-0.89$ \\
\ion{Sc}{2} & $4320.732$ & $0.61$ & $-0.25$ & $<10$ & $<-0.75$ \\
\ion{Ti}{2} & $3349.408$ & $0.05$ & $ 0.59$ & $ 47$ & $-0.78$ \\
\ion{Ti}{2} & $3361.218$ & $0.03$ & $ 0.28$ & $ 33$ & $-0.75$ \\
\ion{Ti}{2} & $3372.800$ & $0.01$ & $ 0.27$ & $ 47$ & $-0.50$ \\
\ion{Ti}{2} & $3759.296$ & $0.61$ & $ 0.27$ & $ 23$ & $-0.41$ \\
\ion{Ti}{2} & $3761.323$ & $0.57$ & $ 0.17$ & $ 13$ & $-0.65$ \\
\ion{Cr}{1} & $4254.332$ & $0.00$ & $-0.11$ & $<10$ & $<0.65$ \\
\ion{Cr}{1} & $4274.796$ & $0.00$ & $-0.23$ & $<10$ & $<0.76$ \\
\ion{Mn}{1} & $4033.062$ & $0.00$ & $-0.62$ & $<10$ & $<0.47$ \\
\ion{Fe}{1} & $3440.606$ & $0.00$ & $-0.67$ & $ 62$ & $ 2.10$ \\
\ion{Fe}{1} & $3440.989$ & $0.05$ & $-0.96$ & $ 60$ & $ 2.41$ \\
\ion{Fe}{1} & $3465.861$ & $0.11$ & $-1.19$ & $ 34$ & $ 2.20$ \\
\ion{Fe}{1} & $3475.450$ & $0.09$ & $-1.05$ & $ 35$ & $ 2.07$ \\
\ion{Fe}{1} & $3490.574$ & $0.05$ & $-1.11$ & $ 32$ & $ 2.02$ \\
\ion{Fe}{1} & $3565.379$ & $0.96$ & $-0.13$ & $ 27$ & $ 1.95$ \\
\ion{Fe}{1} & $3570.098$ & $0.92$ & $ 0.15$ & $ 39$ & $ 1.84$ \\
\ion{Fe}{1} & $3581.193$ & $0.86$ & $ 0.41$ & $ 49$ & $ 1.71$ \\
\ion{Fe}{1} & $3608.859$ & $1.01$ & $-0.10$ & $ 29$ & $ 2.00$ \\
\ion{Fe}{1} & $3618.768$ & $0.99$ & $ 0.00$ & $ 30$ & $ 1.90$ \\
\ion{Fe}{1} & $3727.619$ & $0.96$ & $-0.63$ & $ 17$ & $ 2.11$ \\
\ion{Fe}{1} & $3758.233$ & $0.96$ & $-0.03$ & $ 39$ & $ 1.99$ \\
\ion{Fe}{1} & $3763.789$ & $0.99$ & $-0.24$ & $ 23$ & $ 1.93$ \\
\ion{Fe}{1} & $3815.840$ & $1.49$ & $ 0.24$ & $ 24$ & $ 2.02$ \\
\ion{Fe}{1} & $3820.425$ & $0.86$ & $ 0.12$ & $ 49$ & $ 1.91$ \\
\ion{Fe}{1} & $3824.444$ & $0.00$ & $-1.36$ & $ 42$ & $ 2.30$ \\
\ion{Fe}{1} & $3825.881$ & $0.92$ & $-0.04$ & $ 41$ & $ 1.99$ \\
\ion{Fe}{1} & $3840.438$ & $0.99$ & $-0.51$ & $ 21$ & $ 2.13$ \\
\ion{Fe}{1} & $3856.372$ & $0.05$ & $-1.29$ & $ 38$ & $ 2.21$ \\
\ion{Fe}{1} & $3859.911$ & $0.00$ & $-0.71$ & $ 67$ & $ 2.08$ \\
\ion{Fe}{1} & $3922.912$ & $0.05$ & $-1.65$ & $ 22$ & $ 2.24$ \\
\ion{Fe}{1} & $4045.812$ & $1.49$ & $ 0.28$ & $ 28$ & $ 2.03$ \\
\ion{Fe}{1} & $4063.594$ & $1.56$ & $ 0.06$ & $ 15$ & $ 2.00$ \\
\ion{Fe}{1} & $4071.738$ & $1.61$ & $-0.02$ & $ 16$ & $ 2.17$ \\
\ion{Fe}{1} & $5269.538$ & $0.86$ & $-1.32$ & $  8$ & $ 2.20$ \\
\ion{Fe}{2} & $5018.440$ & $2.89$ & $-1.22$ & $<10$ & $<3.00$ \\
\ion{Fe}{2} & $5169.033$ & $2.89$ & $-1.30$ & $<10$ & $<3.08$ \\
\ion{Co}{1} & $3405.114$ & $0.43$ & $ 0.25$ & $<39$ & $<1.08$ \\
\ion{Co}{1} & $3453.508$ & $0.43$ & $ 0.38$ & $<35$ & $<0.86$ \\
\ion{Co}{1} & $3483.405$ & $0.51$ & $-1.00$ & $<32$ & $<2.27$ \\
\ion{Co}{1} & $3502.617$ & $0.17$ & $-1.24$ & $<32$ & $<2.12$ \\
\ion{Co}{1} & $3873.114$ & $0.43$ & $-0.66$ & $<19$ & $<1.44$ \\
\ion{Ni}{1} & $3414.760$ & $0.03$ & $-0.01$ & $ 32$ & $ 0.58$ \\
\ion{Ni}{1} & $3461.649$ & $0.03$ & $-0.35$ & $ 20$ & $ 0.63$ \\
\ion{Ni}{1} & $3515.049$ & $0.11$ & $-0.21$ & $ 21$ & $ 0.59$ \\
\ion{Ni}{1} & $3524.535$ & $0.03$ & $ 0.01$ & $ 36$ & $ 0.60$ \\
\ion{Zn}{1} & $3345.015$ & $4.08$ & $ 0.25$ & $<48$ & $<2.72$ \\
\ion{Zn}{1} & $3345.570$ & $4.08$ & $-0.50$ & $<48$ & $<3.46$ \\
\ion{Zn}{1} & $3345.937$ & $4.08$ & $-1.68$ & $<48$ & $<4.64$ \\
\ion{Zn}{1} & $4810.528$ & $4.08$ & $-0.14$ & $<10$ & $<1.97$ \\
\ion{Sr}{2} & $4077.709$ & $0.00$ & $ 0.17$ & $<10$ & $<-2.83$ \\
\ion{Sr}{2} & $4215.519$ & $0.00$ & $-0.14$ & $<10$ & $<-2.53$ \\
\ion{Ba}{2} & $4934.076$ & $0.00$ & $-0.15$ & $<10$ & $<-2.33$ \\
\ion{Eu}{2} & $4129.725$ & $0.00$ & $ 0.17$ & $<10$ & $<-1.99$ 
\enddata
\end{deluxetable} 

\begin{deluxetable}{rrcrrrr} 
\tablecolumns{7} 
\tablewidth{0pt} 
\tablecaption{\label{Tab:MARCSmodel} Structure of the MARCS model used in 
   the abundance analysis} 
\tablehead{
  \colhead{$\log \tau_{\mbox{\scriptsize Ross}}$} &
  \colhead{$\log\tau_{5000}$} & \colhead{$T$} & 
  \colhead{$\log P_g$} & \colhead{$\log P_e$} & \colhead{$\log \kappa_{\mbox{\scriptsize Ross}}$} &
  \colhead{$\log \rho$}\\
  \colhead{} & \colhead{} & \colhead{(K)} & \colhead{(dyn/cm$^2$)} & \colhead{(dyn/cm$^2$)} & 
  \colhead{(cm$^2$/g)} & \colhead{(g\,/cm$^{3}$)}
}
\startdata 
 $-4.0$ & $-3.215$ & $4037$ & $2.31$ & $-3.04$ & $-4.06$ & $-9.04$\\
 $-3.8$ & $-3.025$ & $4062$ & $2.46$ & $-2.95$ & $-3.99$ & $-8.97$\\
 $-3.6$ & $-2.855$ & $4092$ & $2.60$ & $-2.81$ & $-3.88$ & $-8.83$\\
 $-3.4$ & $-2.700$ & $4120$ & $2.73$ & $-2.69$ & $-3.78$ & $-8.71$\\
 $-3.2$ & $-2.551$ & $4144$ & $2.85$ & $-2.57$ & $-3.69$ & $-8.59$\\
 $-3.0$ & $-2.404$ & $4162$ & $2.97$ & $-2.48$ & $-3.61$ & $-8.47$\\
 $-2.8$ & $-2.256$ & $4175$ & $3.09$ & $-2.39$ & $-3.53$ & $-8.35$\\
 $-2.6$ & $-2.104$ & $4186$ & $3.22$ & $-2.30$ & $-3.46$ & $-8.23$\\
 $-2.4$ & $-1.950$ & $4199$ & $3.34$ & $-2.22$ & $-3.39$ & $-8.11$\\
 $-2.2$ & $-1.793$ & $4216$ & $3.46$ & $-2.12$ & $-3.30$ & $-7.99$\\
 $-2.0$ & $-1.634$ & $4237$ & $3.58$ & $-2.02$ & $-3.21$ & $-7.87$\\
 $-1.8$ & $-1.474$ & $4265$ & $3.70$ & $-1.90$ & $-3.11$ & $-7.75$\\
 $-1.6$ & $-1.313$ & $4301$ & $3.81$ & $-1.78$ & $-3.00$ & $-7.65$\\
 $-1.4$ & $-1.150$ & $4348$ & $3.91$ & $-1.63$ & $-2.88$ & $-7.55$\\
 $-1.2$ & $-0.986$ & $4408$ & $4.01$ & $-1.47$ & $-2.74$ & $-7.45$\\
 $-1.0$ & $-0.820$ & $4495$ & $4.10$ & $-1.26$ & $-2.57$ & $-7.37$\\
 $-0.9$ & $-0.737$ & $4538$ & $4.14$ & $-1.16$ & $-2.48$ & $-7.34$\\
 $-0.8$ & $-0.653$ & $4596$ & $4.18$ & $-1.04$ & $-2.38$ & $-7.30$\\
 $-0.7$ & $-0.568$ & $4662$ & $4.21$ & $-0.91$ & $-2.28$ & $-7.28$\\
 $-0.6$ & $-0.482$ & $4738$ & $4.25$ & $-0.77$ & $-2.16$ & $-7.25$\\
 $-0.5$ & $-0.396$ & $4825$ & $4.27$ & $-0.62$ & $-2.04$ & $-7.23$\\
 $-0.4$ & $-0.309$ & $4927$ & $4.30$ & $-0.45$ & $-1.90$ & $-7.22$\\
 $-0.3$ & $-0.220$ & $5040$ & $4.32$ & $-0.27$ & $-1.76$ & $-7.20$\\
 $-0.2$ & $-0.131$ & $5174$ & $4.34$ & $-0.07$ & $-1.61$ & $-7.20$\\
 $-0.1$ & $-0.041$ & $5321$ & $4.36$ & $ 0.14$ & $-1.45$ & $-7.19$\\
 $ 0.0$ & $ 0.051$ & $5495$ & $4.37$ & $ 0.37$ & $-1.27$ & $-7.19$\\
 $ 0.1$ & $ 0.143$ & $5686$ & $4.38$ & $ 0.60$ & $-1.09$ & $-7.20$\\
 $ 0.2$ & $ 0.235$ & $5917$ & $4.39$ & $ 0.86$ & $-0.89$ & $-7.20$\\
 $ 0.3$ & $ 0.327$ & $6189$ & $4.40$ & $ 1.15$ & $-0.67$ & $-7.22$\\
 $ 0.4$ & $ 0.419$ & $6443$ & $4.41$ & $ 1.39$ & $-0.47$ & $-7.23$\\
 $ 0.5$ & $ 0.509$ & $6663$ & $4.41$ & $ 1.59$ & $-0.31$ & $-7.24$\\
 $ 0.6$ & $ 0.599$ & $6854$ & $4.42$ & $ 1.75$ & $-0.17$ & $-7.24$\\
 $ 0.8$ & $ 0.778$ & $7158$ & $4.42$ & $ 1.99$ & $ 0.03$ & $-7.26$\\
 $ 1.0$ & $ 0.957$ & $7420$ & $4.43$ & $ 2.18$ & $ 0.21$ & $-7.26$\\
 $ 1.2$ & $ 1.137$ & $7656$ & $4.44$ & $ 2.34$ & $ 0.36$ & $-7.27$\\
 $ 1.4$ & $ 1.318$ & $7877$ & $4.45$ & $ 2.49$ & $ 0.50$ & $-7.27$\\
 $ 1.6$ & $ 1.501$ & $8090$ & $4.46$ & $ 2.62$ & $ 0.63$ & $-7.28$\\
 $ 1.8$ & $ 1.686$ & $8301$ & $4.47$ & $ 2.74$ & $ 0.76$ & $-7.28$\\
 $ 2.0$ & $ 1.872$ & $8513$ & $4.48$ & $ 2.86$ & $ 0.88$ & $-7.28$
\enddata
\end{deluxetable} 

\begin{deluxetable}{lrrrrrrl} 
\tablecolumns{8} 
\tablewidth{0pc} 
\tablecaption{\label{Tab:Abundances} Abundances and relative abundances of HE~0107$-$5240} 
\tablehead{ 
\colhead{El.} & \colhead{Ion} & \colhead{$N_{\mbox{\scriptsize lines}}$} & 
\colhead{$\log\epsilon (\mbox{X})$} & 
\colhead{$\log\epsilon (\mbox{X})_{\odot}$\tablenotemark{a}} & \colhead{[X/H]} &
\colhead{[X/Fe]} & \colhead{Notes}}
\startdata 
  Li &       1 &       2 & $<1.12$ & 1.10 & $<0.02$ & $<5.3$ &          \\
  C  & \nodata & \nodata &  7.11 & 8.39 & $-1.28$ &  $+4.00$ & Synth. of $C_2$   \\
  C  & \nodata & \nodata &  6.81 & 8.39 & $-1.58$ &  $+3.70$ & Synth. of CH-AX   \\
  N  & \nodata & \nodata &  4.93 & 7.93 & $-3.00$ &  $+2.28$ & Synth. of CN, assuming C=7.11 \\
  N  & \nodata & \nodata &  5.22 & 7.93 & $-2.71$ &  $+2.57$ & Synth. of CN, assuming C=6.81 \\
  Na &  1 &    2 &     1.86 & 6.33 &  $-4.47$ &  $+0.81$ &                            \\
  Mg &  1 &    2 &     2.41 & 7.54 &  $-5.13$ &  $+0.15$ &                            \\
  Al &  1 &    1 &  $<0.93$ & 6.47 & $<-5.54$ & $<-0.26$ & $\lambda = 3961.52$\,{\AA} \\
  Si &  1 &    1 &  $<2.55$ & 7.51 & $<-4.96$ & $<+0.32$ &                            \\
  S  &  1 &    1 &  $<7.11$ & 7.33 & $<-0.22$ & $<+5.06$ & $\lambda = 4034.03$\,{\AA} \\
  Ca &  1 &    1 &     0.99 & 6.36 &  $-5.37$ &  $-0.09$ &                            \\
  Ca &  2 &    2 &     1.44 & 6.36 &  $-4.92$ &  $+0.36$ &                            \\
  Sc &  2 &    1 & $<-1.50$ & 3.17 & $<-4.67$ & $<+0.61$ & $\lambda = 3613.83$\,{\AA} \\
  Ti &  2 &    5 &  $-0.62$ & 5.02 &  $-5.64$ &  $-0.36$ &                            \\
  Cr &  1 &    1 &  $<0.65$ & 5.67 & $<-5.02$ & $<+0.26$ & $\lambda = 4254.33$\,{\AA} \\
  Mn &  1 &    1 &  $<0.47$ & 5.39 & $<-4.92$ & $<+0.36$ & $\lambda = 4033.06$\,{\AA} \\
  Fe &  1 &   25 &     2.06 & 7.45 &  $-5.39$ & \nodata  & LTE                        \\
  Fe &  1 & \nodata &  2.17 & 7.45 &  $-5.28$ & \nodata  & non-LTE                    \\
  Fe &  2 &    1 &  $<3.00$ & 7.45 & $<-4.45$ & \nodata  & $\lambda = 5018.44$\,{\AA} \\
  Co &  1 &    1 &  $<0.86$ & 4.92 & $<-4.06$ & $<+1.22$ & $\lambda = 3453.51$\,{\AA}  \\
  Ni &  1 &    4 &     0.60 & 6.25 &  $-5.65$ &  $-0.37$ &                             \\
  Zn &  1 &    1 &  $<1.97$ & 4.60 & $<-2.63$ & $<+2.65$ & $\lambda = 4810.53$\,{\AA}  \\
  Sr &  2 &    1 & $<-2.83$ & 2.97 & $<-5.80$ & $<-0.52$ & $\lambda = 4077.71$\,{\AA}  \\
  Ba &  2 &    1 & $<-2.33$ & 2.13 & $<-4.46$ & $<+0.82$ & $\lambda = 4934.08$\,{\AA}  \\
  Eu &  2 &    1 & $<-1.99$ & 0.51 & $<-2.50$ & $<+2.78$ & $\lambda = 4129.73$\,{\AA}  \\
\enddata 
\tablenotetext{a}{See \S \ref{Sect:SolarAbundances} for sources.}
\end{deluxetable} 

\begin{deluxetable}{lrrrrrcl} 
\tablecolumns{8} 
\tablewidth{0pc} 
\tablecaption{\label{Tab:AbundanceChanges} Sensitivity of abundances to changes of 
  stellar parameters and equivalent widths.} 
\tablehead{
  \colhead{El.} & \colhead{Ion} & \colhead{{\tefft}} & \colhead{$\log g$}
  & \colhead{$v_{\mbox{\scriptsize micr}}$} & \colhead{$W_{\lambda}$} &
  \colhead{$\sigma\left(\teffm + \log g + v_{\mbox{\scriptsize micr}}\right)$}& \colhead{Notes}\\
  \colhead{} & \colhead{} & \colhead{($+150$\,K)} & \colhead{($+0.3$\,dex)}
  & \colhead{($+0.5$\,km\,s$^{-1}$)} & \colhead{($+3$\,m{\AA})} & \colhead{(dex)} & \colhead{} 
  }
\startdata 
  C  & \nodata & $+0.23$ & $-0.07$ & $+0.01$ & \nodata & $0.24$ & Synth. of $C_2$\\
  C  & \nodata & $+0.32$ & $-0.11$ & $-0.02$ & \nodata & $0.34$ & Synth. of CH-AX\\
  N  & \nodata & $+0.17$ & $-0.03$ & $-0.02$ & \nodata & $0.17$ & Synth. of CN; C from $C_2$\\
  N  & \nodata & $+0.09$ & $+0.02$ & $+0.03$ & \nodata & $0.10$ & Synth. of CN; C from CH\\
  Na &  1      & $+0.14$ & $+0.00$ & $-0.01$ & $+0.07$ & $0.15$ & \\
  Mg &  1      & $+0.12$ & $+0.00$ & $ 0.00$ & $+0.10$ & $0.14$ & \\
  Al &  1      & $+0.14$ & $-0.01$ & $-0.01$ & $+0.12$ & $0.18$ & \\
  Si &  1      & $+0.15$ & $+0.01$ & $-0.01$ & $+0.07$ & $0.17$ & \\
  Ca &  1      & $+0.15$ & $-0.01$ & $-0.02$ & $+0.06$ & $0.16$ & \\
  Ca &  2      & $+0.17$ & $-0.02$ & $-0.08$ & $+0.02$ & $0.19$ & \\
  Sc &  2      & $+0.12$ & $+0.09$ & $-0.02$ & $+0.07$ & $0.17$ & \\
  Ti &  2      & $+0.11$ & $+0.09$ & $-0.03$ & $+0.07$ & $0.15$ & \\
  Cr &  1      & $+0.17$ & $-0.01$ & $-0.01$ & $+0.12$ & $0.21$ & \\
  Mn &  1      & $+0.20$ & $ 0.00$ & $ 0.00$ & $+0.13$ & $0.24$ & \\
  Fe &  1      & $+0.20$ & $-0.01$ & $-0.04$ & $+0.07$ & $0.20$ & \\
  Fe &  2      & $+0.06$ & $+0.10$ & $ 0.00$ & $+0.13$ & $0.15$ & \\
  Co &  1      & $+0.21$ & $-0.01$ & $-0.04$ & $+0.06$ & $0.22$ & \\
  Ni &  1      & $+0.21$ & $-0.01$ & $-0.03$ & $+0.07$ & $0.21$ & \\
  Zn &  1      & $+0.08$ & $+0.04$ & $-0.01$ & $+0.13$ & $0.16$ & \\
  Sr &  2      & $+0.11$ & $+0.10$ & $-0.01$ & $+0.13$ & $0.20$ & \\
  Ba &  2      & $+0.12$ & $+0.09$ & $-0.01$ & $+0.12$ & $0.19$ & \\
  Eu &  2      & $+0.11$ & $+0.09$ & $-0.01$ & $+0.13$ & $0.19$ & \\
\enddata 
\end{deluxetable} 
\clearpage

\end{document}